\documentclass[12pt]{article}
\pdfoutput=1
\usepackage{titling}
\usepackage{amsmath}
\usepackage{slashed}
\usepackage{amssymb}
\usepackage{epsfig}
\usepackage{graphicx}
\usepackage{multirow}
\usepackage{color}
\usepackage[normal]{subfigure}
\usepackage{rotating}
\usepackage{hyperref}
\usepackage[margin=0.9in]{geometry}
\usepackage[table]{xcolor}
\usepackage{enumitem}
\usepackage[utf8x]{inputenc}
\usepackage[compress,numbers,sort]{natbib}
\usepackage{authblk}
\usepackage{colortbl}
\usepackage{pdflscape}
\usepackage{color}

\usepackage{caption}
\definecolor{nicered}{rgb}{0.7,0.1,0.1}
\definecolor{nicegreen}{rgb}{0.1,0.5,0.1}
\definecolor{red}{rgb}{1.0, 0, 0}
\hypersetup{colorlinks,citecolor= nicegreen,linkcolor= nicered}

\def\eq#1{{Eq.~(\ref{#1})}}
\def\eqs#1#2{{Eqs.~(\ref{#1})--(\ref{#2})}}
\def\fig#1{{Fig.~\ref{#1}}}

\def\Table#1{{Table~\ref{#1}}}

\def\sect#1{{Sect.~\ref{#1}}}

\def\app#1{{Appendix~\ref{#1}}}

\def\abs#1{\left| #1\right|}

\def\Im{\mbox{Im}\,}
\def\Re{\mbox{Re}\,}
\def\Tr{\mbox{Tr}\,}

\renewcommand{\bar}{\overline}

\definecolor{LightCyan}{rgb}{0.88,1,1}
\definecolor{piggypink}{rgb}{0.99, 0.87, 0.9}
\definecolor{applegreen}{rgb}{0.55, 0.71, 0.0}
\definecolor{darkpastelgreen}{rgb}{0.01, 0.75, 0.24}
\definecolor{green-yellow}{rgb}{0.68, 1.0, 0.18}

\newcommand{\beq}{\begin{equation}}
\newcommand{\eeq}{\end{equation}}
\newcommand{\bea}{\begin{eqnarray}}
\newcommand{\eea}{\end{eqnarray}}

\title{\bf{\huge Implications of perturbative unitarity for scalar di-boson resonance searches at LHC}}

\author[1,2]{\large Luca Di Luzio\thanks{luca.di-luzio@durham.ac.uk}}
\author[3,4]{\large Jernej F. Kamenik\thanks{jernej.kamenik@cern.ch}}
\author[5,6]{\large Marco Nardecchia\thanks{marco.nardecchia@cern.ch}}

\affil[1]{\emph{\normalsize Dipartimento di Fisica, Universit\`a di Genova and INFN, Sezione di Genova, 
\newline via Dodecaneso 33, 16159 Genova, Italy}}
\affil[2]{\emph{\normalsize Institute for Particle Physics Phenomenology, Department of Physics, Durham University, DH1 3LE, United Kingdom}}
\affil[3]{\emph{\normalsize Jo\v{z}ef Stefan Institute, Jamova 39, 1000 Ljubljana, Slovenia}}
\affil[4]{\emph{\normalsize Faculty of Mathematics and Physics, University of Ljubljana, Jadranska 19, 1000 Ljubljana, \newline Slovenia}}
\affil[5]{\emph{\normalsize DAMTP, University of Cambridge, Wilberforce Road, Cambridge CB3 0WA, \newline United Kingdom}}
\affil[6]{\emph{\normalsize CERN, Theoretical Physics Department, Geneva, Switzerland}}

\date{}

\begin{document} 

\maketitle

\begin{abstract}
\normalsize

We study the constraints implied by partial wave unitarity on new physics in the form of spin-zero di-boson resonances at LHC. 
We derive the scale where the effective description in terms of the SM supplemented 
by a single resonance is expected to break down depending on the resonance mass and signal cross-section. 
Likewise, we use unitarity arguments in order to set perturbativity bounds on renormalizable UV completions 
of the effective description. We finally discuss under which conditions scalar di-boson resonance signals can be accommodated within weakly-coupled models.

\end{abstract}

\clearpage

\tableofcontents

\clearpage

\section{Introduction}
\label{intro}

Unitarity of the time evolution of an isolated quantum system and in particular of the associated $S$-matrix is one of the cornerstones of quantum field theory. In practical perturbative calculations however, $S$-matrix unitarity is always approximate and asymptotic. Nonetheless, significant violations of unitarity at low orders in perturbation theory are heralds of a strongly-coupled system and can be used to constrain the range of validity of a given (effective) quantum field theory description. 

Perhaps most famously, constraints imposed by perturbative unitarity in $WW$ scattering have been used in the past to infer an upper bound on the Higgs boson mass or, alternatively, on the scale where the standard model (SM) description of weak interactions would need to be completed in the ultraviolet (UV) in terms of some new strongly coupled dynamics \cite{Lee:1977yc,Lee:1977eg}. Correspondingly it allowed to narrow down the relevant mass search window and motivate the construction of the LHC with capabilities that ensured the eventual Higgs boson discovery (cf.~\cite{Djouadi:2005gi} for a review).

More generally, perturbative unitarity constraints on the validity of a certain theoretical description are applicable both in non-renormalizable as well as renormalizable models. In both cases they allow to assess the limitations of a perturbative expansion. In the non-renormalizable effective field theory (EFT) approach this amounts to a truncated power expansion in $(E/\Lambda)$, where $E$ is a typical energy in a process and $\Lambda$ is the EFT cut-off scale. Violations of perturbative unitarity signal the breakdown of such an expansion, when the leading powers do not represent a good approximation to the physical result. A notable standard example is the pion scattering in chiral perturbation theory, where the loop and power expansion are adequate at low enough scattering energies but violate perturbative unitarity at higher energies and eventually need to be UV completed with the inclusion of dynamical vector resonances.  On the other hand within renormalizable models, the expansion proceeds in terms of positive powers of the renormalizable couplings. Sizable violations of unitarity at leading (tree) order signal the breakdown of such an expansion and the onset of strongly coupled dynamics. Here the most renown case is that of the aforementioned $WW$ boson scattering in presence of a heavy SM Higgs boson.


The recently rekindled interest in new physics (NP) in the form of (possibly broad) di-photon resonances~\cite{ATLAS-CONF-2015-081, CMS:2015dxe, CMS:2016owr, ATLAS:Moriond, CMS:Moriond} at the LHC prompt us to reconsider the implications of perturbative unitarity for EFT interpretations of resonances decaying to di-boson final states. In particular, focusing on promptly produced scalar SM singlets decaying to two SM gauge bosons we aim to address the following questions: at which maximal energies do we expect the effective description in terms of the SM supplemented by a single scalar to break down? What can we learn about the possible UV completions of such effective theory from unitarity arguments? In particular, whether and under which conditions can a potential di-boson signal be accommodated within weakly-coupled models? 

We further motivate the endeavor with the observation that in perturbative weakly-coupled models, decays of a scalar singlet into two transverse SM gauge bosons can only arise at loop level involving massive charged and/or colored particles leading to a suppression factor of $\Gamma_{V_T V_T}/M \propto \alpha_{V}^2/16\pi^3$. Even in the case of QCD $\Gamma_{V_T V_T}/M \gtrsim 10^{-4}$  would require large couplings and/or large multiplicies of new states contributing in the loop. Both possibilities are potentially subject to constraints coming from perturbative unitarity.  In particular, we will  show how they enter the amplitudes of $2 \to 2$ scatterings of the new degrees of freedom. 

Similar considerations have already triggered several studies addressing the issue of the predictivity and calculability within weakly-coupled perturbative models of di-photon resonances.\footnote{For a broad survey of such models cf.~\cite{Staub:2016dxq}.} These include studying the renormalization group equations (RGE) of the models~\cite{Goertz:2015nkp} or the actual appearance of Landau poles~\cite{Son:2015vfl,Franceschini:2015kwy,Cao:2015twy}. For marginal operators such as those corresponding to the gauge couplings, Yukawas or the scalar quartic, both effects are however only logarithmically sensitive to the UV cut-off scale of the theory. The resulting constraints can also be circumvented if the models can be UV completed into theories exhibiting an infrared (IR) fixed point behavior. In case of scalar extensions, the stability of the scalar potential has also been used~\cite{Salvio:2016hnf,Ge:2016xcq}. In this case the possibility of a metastable vacuum with its intricate relations to the cosmological history of the Universe requires additional assumptions going beyond quantum field theory arguments. Some aspects of partial wave unitarity for di-photon resonances which partially overlap with our work 
have already been discussed in~\cite{Murphy:2015kag,Fabbrichesi:2016alj,Cynolter:2016jxv}, however with a different focus with respect to our analysis.

The rest of the paper is structured as follows: \sect{reviewPWU} contains a brief recap of partial wave unitarity arguments, which we first apply in \sect{EFT} to the EFT 
case where a di-boson resonance is the only new degree of freedom beyond the SM. In \sect{wcmodels} 
we then consider weakly-coupled benchmark models with either new fermionic or scalar degrees of freedom coupling to a di-boson resonance and inducing 
the EFT operators in the low-energy limit. Our main results 
are summarized in~\sect{concl}. Finally, some relevant technical details of our computations can be found in \app{Amplitudes}. 

\section{Brief review on partial wave unitarity}
\label{reviewPWU}

Let us denote by $\mathcal{T}_{fi} (\sqrt{s},\cos\theta)$ the matrix element of a $2\to 2$ scattering amplitude in momentum space, defined via  
\beq
\label{defT}
(2\pi)^4 \delta^{(4)} (P_i - P_f) \mathcal{T}_{fi} (\sqrt{s},\cos\theta) = \langle f | T | i \rangle \, ,
\eeq
where $T$ is the interacting part of the $S$-matrix, $S = 1+ i T$.  
The dependence of the scattering amplitude on $\cos\theta$ is eliminated 
by projecting it onto partial waves of total angular momentum $J$ (see e.g.~\cite{Itzykson:1980rh,Chanowitz:1978mv,Schuessler:2007av})
\beq 
\label{PWprojprev}
a^J_{fi} = \frac{\beta_f^{1/4}(s,m^2_{f1},m^2_{f2}) \beta_i^{1/4}(s,m^2_{i1},m^2_{i2})}{32 \pi s} \int_{-1}^{1}
d(\cos\theta) \, d^J_{\mu_i\mu_f}(\theta) \, \mathcal{T}_{fi} (\sqrt{s},\cos\theta) \, ,
\eeq
where $d^J_{\mu_i\mu_f}$ is the $J$-th Wigner $d$-function appearing in the Jacob-Wick 
expansion \cite{Jacob:1959at}, while $\mu_i = \lambda_{i1} - \lambda_{i2}$ 
and $\mu_f = \lambda_{f1} - \lambda_{f2}$ are defined in terms of the helicities of the 
initial ($\lambda_{i1}, \lambda_{i2}$) and final ($\lambda_{f1}, \lambda_{f2}$) states. 
The function $\beta(x,y,z) = x^2 + y^2 + z^2 - 2xy - 2yz - 2zx$ is a kinematical factor related to the momentum (to the fourth power) of a 
given particle in the center of mass frame. 
The right hand side of \eq{PWprojprev} must be further multiplied by a $\frac{1}{\sqrt{2}}$ factor 
for any identical pair of particles either in the initial or final state. 

\begin{figure}[!ht]
  \begin{center}
    \includegraphics[width=.40\textwidth]{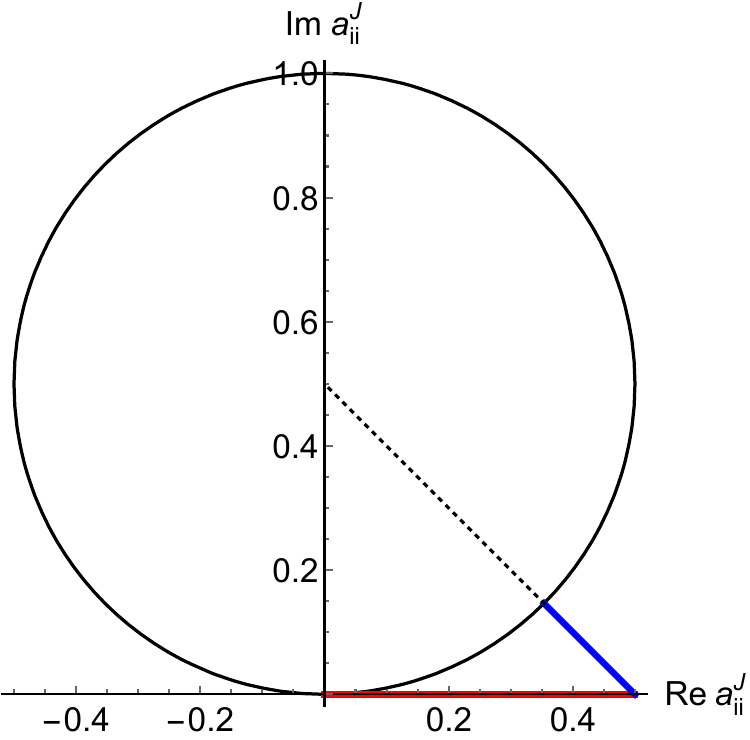}
  \end{center}
  \caption{\label{Argand}
  Unitarity constraint in the Argand plane. A Born value of $\Re a^J_{ii} = \frac{1}{2}$ and $\Im a^J_{ii} = 0$ (red line) requires a correction (blue line) 
  which amounts to at least the $\sqrt{2}-1\simeq 40 \%$ of the the tree-level value in order to come back inside the unitarity circle.}
\end{figure}

When restricted to a same-helicity state (zero total spin), the Wigner $d$-functions reduce 
to the Legendre polynomials, i.e.~$d^J_{00} = P_J$. 
In practice, we will only focus on $J=0$ ($d^0_{00} = P_0 = 1$), 
since higher partial waves typically give smaller amplitudes, 
unless $J=0$ amplitudes are suppressed or vanish for symmetry reasons. 
Hence, the quantity we are interested in is 
\beq 
\label{PWproj}
a^0_{fi} = \frac{\beta_f^{1/4}(s,m^2_{f1},m^2_{f2}) \beta_i^{1/4}(s,m^2_{i1},m^2_{i2})}{32 \pi s} \int_{-1}^{1}
d(\cos\theta) \, \mathcal{T}_{fi} (\sqrt{s},\cos\theta) \, . 
\eeq
In the high-energy limit, $\sqrt{s} \rightarrow \infty$, one has $\beta_f^{1/4}\beta_i^{1/4} / s \to 1$.
The unitarity condition on the $S$-matrix, $SS^\dag=1$, gives 
\beq
\label{unitS}
\frac{1}{2i} \left( a^J_{fi} - a^{J*}_{if} \right) \geq \sum_h a^{J*}_{hf} a^{J}_{hi} \, ,
\eeq
where the sum over $h$ is restricted to 2-particle states, which slightly underestimates the left hand side. 
For $i=f$ \eq{unitS} reduces to 
\beq 
\label{UnitBoundprev}
\Im a^J_{ii} \geq |a^J_{ii}|^2 \, .
\eeq
Hence, $a^J_{ii}$ must lie inside the circle in the Argand plane defined by (cf.~also \fig{Argand})
\beq 
\label{UnitBoundprev2}
\left(\Re a^J_{ii}\right)^2 + \left(\Im a^J_{ii} - \frac{1}{2} \right)^2 \leq \frac{1}{4} \, , 
\eeq
which implies 
\beq
\label{UnitBoundprev3}
|\Im a^J_{ii} | \leq 1 \qquad \text{and} \qquad |\Re a^J_{ii} | \leq \frac{1}{2} \, .
\eeq
Under the assumption that the tree-level amplitude is real, 
\eq{UnitBoundprev3} suggests the following perturbativity criterium
\beq
\label{UnitBound}
|\Re (a^J_{ii})^{\text{Born}} | \leq \frac{1}{2} \, .
\eeq
In fact, a Born value of $\Re a^J_{ii} = \frac{1}{2}$ and $\Im a^J_{ii} = 0$ 
needs at least a correction of $40\%$ in order to restore unitarity (cf.~\fig{Argand}). 

In reality, one expects to have issues with perturbativity even before saturating the bound in \eq{UnitBound}, 
which is hence understood to be a conservative one. 
Stronger constraints can be obtained by considering the full transition matrix 
connecting all the possible 2-particle states, 
which amount to applying \eq{UnitBound} to the highest eigenvalue of $|\Re (a^J_{if})^{\text{Born}} |$.


\section{Effective field theory of a scalar resonance}
\label{EFT}

We consider the EFT of a gauge singlet spin-0 resonance, $S$ with mass $M_S$, coupled to the SM 
fields. Assuming CP invariance, we choose $S$ to transform as a 
scalar.\footnote{The pseudo-scalar case leads to analogous conclusions as far as unitarity bounds are concerned, 
hence in the following we will not consider it separately.} 
The only renormalizable terms couple $S$ to the Higgs in the scalar potential 
\beq 
\mathcal{L}^{(4)}_{\text{int.}} = - \mu_S S H^\dag H - \frac{\lambda_S}{2} S^2 H^\dag H \, , 
\label{eq:L4}
\eeq
where $\mu_S \lesssim s_\alpha m_S^2/v \lesssim m_S^2/600$\,GeV. In the inequality we have introduced $v\simeq 246$~GeV and $s_\alpha\lesssim 0.4$~\cite{Falkowski:2015swt, ATLASCONF2015044} as the sine of the mixing angle between $S$ and the Higgs, 
$h$ (in the unitary gauge $H = (0, v+h)/\sqrt{2}$). While for a CP-even $S$ the $ \mu_S$ term can contribute to the $S \to hh,W_LW_L,~Z_LZ_L$ widths, it is not relevant for unitarity bounds in the high-energy limit. The $d=5$ Lagrangian instead reads
\begin{align}
\label{effLSM}
\mathcal{L}^{(5)}_{\text{int.}} = 
&- \frac{g_3^2}{2 \Lambda_g} S G^2_{\mu\nu}
- \frac{g_2^2}{2 \Lambda_W} S W^2_{\mu\nu} 
- \frac{g_1^2}{2 \Lambda_B} S B^2_{\mu\nu} 
- \frac{1}{\Lambda_H} S \left( D_\mu H \right)^\dag D^\mu H
- \frac{1}{\Lambda'_H} S \left( H^\dag H \right)^2 \nonumber \\
&- \frac{1}{\Lambda_d} S \overline{Q}_L d_R H 
- \frac{1}{\Lambda_u} S \overline{Q}_L u_R H^c 
- \frac{1}{\Lambda_e} S \overline{L}_L e_R H + \text{h.c.} \, ,  
\end{align}
where we have suppressed flavor indices. 
This parametrization makes it clear that apart from the $\mu_S$ term in Eq.~\eqref{eq:L4}, the interactions of a scalar singlet with the SM fields, directly relevant for di-boson resonances at the LHC, are all due to non-renormalizable $d=5$ operators. Their effects are thus expected to be enhanced at high energies eventually leading to the breakdown of perturbative unitarity. In order to quantify this simple observation in the following subsections we evaluate the relevant scattering amplitudes involving SM gauge bosons, Higgs and quarks at the respective leading orders in perturbation theory.
Moreover, since we are interested in studying $2 \to 2$ scattering processes 
at energies $\sqrt{s} \gg M_S \gg v$, we can safely set all the massive parameters (including $M_S$) to zero and work within the unbroken SM theory. 
This also implies that we can neglect any $h$-$S$ mixing effects and set the masses of the final state SM particles to zero. 
We distinguish between two classes of tree-level processes characterized by a different energy scaling of the amplitude: 
scalar mediated scatterings and $d=5$ contact interactions.

\subsection{Scalar mediated boson scattering}
\label{gggamgaminit}

Let us start, as an example, by considering the $\gamma\gamma \to \gamma\gamma$ scattering amplitude 
due to the effective operator 
\beq 
- \frac{e^2}{2 \Lambda_\gamma} S F^2_{\mu\nu} \, ,
\eeq
whose matching with the operators in \eq{effLSM} is given by 
\beq
\frac{1}{\Lambda_\gamma} = \frac{1}{\Lambda_B} + \frac{1}{\Lambda_W} \, .
\eeq
The calculation is detailed in \app{EFTscattering}.  
In the ($++,--$) helicity basis we find
\beq
\label{TEFT}
\mathcal{T} = -\frac{e^4}{\Lambda^2_{\gamma}}
\left( 
\begin{array}{cc}
\frac{s^2}{s-M_S^2} & \frac{s^2}{s-M_S^2} + \frac{t^2}{t-M_S^2} + \frac{u^2}{u-M_S^2}  \\
\frac{s^2}{s-M_S^2} + \frac{t^2}{t-M_S^2} + \frac{u^2}{u-M_S^2}  & \frac{s^2}{s-M_S^2} 
\end{array}
\right) 
\overset{\sqrt{s} \ \gg \ M_S}{\simeq}
-\frac{e^4s}{\Lambda^2_{\gamma}} 
\left( 
\begin{array}{cc}
1 & 0 \\
0 & 1
\end{array}
\right) 
\, ,
\eeq
where 
in the last step we took the high-energy limit. Note that only the $s$-channel survives at high energies. 

The projection on the $J=0$ partial waves is obtained by applying \eq{PWproj} 
and by multiplying by a $1/2$ factor which takes into account the presence of identical particles both in the initial and final states.
In the high-energy limit 
we get
\beq
a^0 \simeq -  
\frac{e^4 s}{32 \pi \Lambda^2_{\gamma}} 
\left( 
\begin{array}{cc}
1 & 0 \\
0 & 1
\end{array}
\right) 
\, , 
\eeq
which, confronted with \eq{UnitBound}, leads to the tree-level unitarity bound
\beq
\sqrt{s} \lesssim \sqrt{16 \pi} \frac{\Lambda_{\gamma}}{e^2} 
\, .
\eeq 
As a matter of fact, the bound above can be made stronger if one considers the full $VV \to V'V'$ scattering matrix, 
where $V$ and $V'$ are any of the $8+3+1$ (transversely polarized) SM gauge bosons of the effective Lagrangian in \eq{effLSM}.  
In such a case, the previous calculation is readily generalized in the high-energy limit where only the $s$-channel survives. 
To this end, we note that a scattering amplitude in the $s$-channel can be written as 
\beq 
\label{factmij}
m_{ij}= \frac{a_i a_j}{s-M^2_S} \, ,
\eeq 
where $a_i$ and $a_j$ are obtained by cutting any $i \to j$ diagram in two parts along the $s$-channel propagator.
The matrix in \eq{factmij} has rank 1 and its non-zero eigenvalue is given by the trace. Hence, 
denoting by $\tilde{a}^0$ the eigenvalue of the $VV \to V'V'$ scattering matrix, in the high-energy limit we get
\beq 
\tilde{a}^0 \simeq  -  
\frac{s}{32 \pi} \left( \frac{8 g_3^4}{\Lambda^2_{g}} + \frac{3 g_2^4}{\Lambda^2_{W}} + \frac{g_1^4}{\Lambda^2_{B}} \right) \, .
\eeq
Correspondingly, the tree-level unitarity bound is given by 
\begin{equation}
\label{boundVV}
\frac{s}{32 \pi} \left(8 \frac{g_s^4}{\Lambda^2_g}  + 3 \frac{g_2^4}{\Lambda^2_W} + \frac{g_1^4}{\Lambda^2_B} \right) \lesssim 
\frac{1}{2} \, . 
\end{equation}
We remark that in deriving these bounds we consider only the transverse polarizations of the $W$ and $Z$ gauge bosons. Generally, scattering amplitudes involving longitudinally polarized massive vector bosons can grow as positive powers of $E/m_{W,Z}$ implying apparently stronger dependence on $s$. However, as it can be easily verified (through an explicit calculation of the processes at hand or more generally via 
a clever gauge choice~\cite{Wulzer:2013mza}), the scattering amplitudes involving longitudinally polarized states sourced by the gauge field strengths in Eq.~\eqref{effLSM} are \textit{suppressed} by powers of $m_{W,Z}/E$ and thus do not lead to relevant unitarity constraints at high $s$.

In the $v \to 0$ limit there is just one additional tree-level $s$-channel contribution 
leading to $2 \to 2$ scatterings of SM particles from \eq{effLSM}, that is due to the operator 
\beq
\label{Higgsscattop}
\frac{1}{\Lambda_H} S (D_{\mu} H)^{\dagger} D^{\mu} H \supset 
\frac{S}{\Lambda_H} \partial_{\mu}H_i^{\dagger} \partial^{\mu} H_i 
\eeq
where we have neglected vertices with 4 or 5 particles 
and $H^T = (H_1, H_2)$. 
In the ($|H^\dag_1 H_1\rangle$, $|H^\dag_2 H_2 \rangle$) basis, 
the $J=0$ partial wave matrix at $\sqrt{s} \gg M_S$ is found to be
\begin{equation}
a^0 \simeq -\frac{s}{64 \pi \Lambda_H^2}
\left(
\begin{array}{cc}
1 & 1\\
1 & 1
\end{array}
\right) \, .
\end{equation}
Imposing the unitarity bound on the highest eigenvalue we get 
\beq 
\sqrt{s} \lesssim \sqrt{32 \pi} \, \Lambda_H \, .
\eeq
Note that in the EW broken vacuum the constraint corresponds to scattering of both the physical Higgs bosons as well as the longitudinally polarized massive EW gauge bosons.
Considering thus also $|H^\dag_i H_i\rangle$ as possible initial and final states,  
\eq{boundVV} is generalized into 
\begin{equation}
\label{boundVVgeneral}
\frac{s}{32 \pi} \left(8 \frac{g_s^4}{\Lambda^2_g}  + 3 \frac{g_2^4}{\Lambda^2_W} + \frac{g_1^4}{\Lambda^2_B} + \frac{1}{2\Lambda^2_H} \right) \lesssim 
\frac{1}{2} \, . 
\end{equation}

\subsection{Fermion-scalar contact interactions}

Next we consider the contact interaction 
\beq 
\label{effopbbbar}
- \frac{1}{\Lambda_d} S \, \overline{Q}_L d_R H = \left[ - \frac{1}{\Lambda_d} \delta^b_a \delta^j_i \right] S \, (\overline{Q}_L)^{ai} (d_R)_{b} H_j \, ,
\eeq
where we have explicitly factored out the color and $SU(2)_L$ group structure. 
In this case the leading scattering process is $\overline{Q} d \to S H$. 
By explicitly writing the polarization and gauge indices in the amplitude, one finds
\beq 
\mathcal{T} = - \frac{\delta^b_a \delta^j_i}{2 \Lambda_d} \overline{v}^s(k) \left( 1 + \gamma_5 \right) u^r(p)  \, . 
\eeq
Only the $++$ and $--$ polarizations survive. By explicit evaluation (cf.~\app{psipsibarpsipsibar} for the expression of the spinor polarizations) we get  
\begin{align}
\mathcal{T}_{++} &= \frac{\delta^b_a \delta^j_i}{\Lambda_d} (E + p^3) \overset{\sqrt{s} \ \gg \ M_S}{\simeq} 
\delta^b_a \delta^j_i \frac{\sqrt{s}}{\Lambda_d}  \, , \\
\mathcal{T}_{--} &=  \frac{\delta^b_a \delta^j_i}{\Lambda_d} (E - p^3) \overset{\sqrt{s} \ \gg \ M_S}{\simeq} 0 \, .
\end{align}
At high energies the $J=0$ partial wave is obtained by considering the color singlet channel 
for a state in the linear combination $\frac{1}{\sqrt{2}} \left( | \overline{Q} d \rangle + | S H \rangle \right)$,
which gives 
\beq
a^0 \simeq \frac{1}{16 \pi} \frac{\sqrt{s}}{\Lambda_d} \, . 
\eeq
Correspondingly, the tree-level unitarity bound reads
\beq 
\label{boundbbbarEFT}
\sqrt{s} \lesssim 8 \pi \Lambda_d \, .
\eeq
Similarly, from the other two contact interactions in the last row of \eq{effLSM} we get 
$\sqrt{s} \lesssim 8 \pi \Lambda_u$ and $\sqrt{s} \lesssim 8 \pi \Lambda_e$.

\subsection{Unitarity bounds}

As an exemplification we consider a scalar resonance $S$ with mass $M_S$ and total width $\Gamma_S$ appearing in a di-photon final state at the LHC.\footnote{Analogous analysis can be performed also for other EW gauge boson final states with the slight complication of disentangling the transverse and longitudinal gauge boson polarizations, as they are sourced by different terms in the EFT Lagrangian ($\Lambda_{B,W}$ and $\Lambda_H$, respectively).}
Expanding the effective Lagrangian in \eq{effLSM} around the broken electroweak (EW) vacuum, 
the part relevant for $S$ production at the LHC is 
\beq 
\label{effL}
\mathcal{L}^{(5)}_{\text{int.}} \supset - \frac{g_3^2}{2 \Lambda_g} S G^2_{\mu\nu}
- \frac{e^2}{2 \Lambda_\gamma} S F^2_{\mu\nu} - \sum_q y_{qS} S \overline{q} q \, ,  
\eeq
whose operators give rise to the decay widths
\begin{align}
\label{Ggammagamma}
\Gamma_{\gamma\gamma} &\equiv \Gamma(S \to \gamma\gamma) = \pi \alpha_{\rm EM}^2 \frac{M_S^3}{\Lambda_\gamma^2} \, , \\
\label{Ggg}
\Gamma_{gg} &\equiv \Gamma(S \to gg) = 8 \pi \alpha_s^2 \frac{M_S^3}{\Lambda_g^2} \, , \\ 
\label{Gbbbar}
\Gamma_{q\overline{q}} &\equiv \Gamma(S \to q \overline{q}) = \frac{3}{8\pi} y_{qS}^2 M_S \left({1-\frac{4m_q^2}{M_S^2}}\right)^{3/2} \, . 
\end{align} 
The matching between the operators in \eq{effL} and \eq{effLSM} then yields 
\beq
\label{matchEFT}
\frac{1}{\Lambda_\gamma} = \frac{1}{\Lambda_B} + \frac{1}{\Lambda_W} \, , 
\qquad 
y_{qS} = \frac{v}{\sqrt 2 \Lambda_q} \, .
\eeq
In the narrow width approximation the prompt $S$ production at the LHC can also be fully parametrized in terms of the relevant decay widths
\beq
\sigma(pp \to S) = \frac{1}{M_S s} \left[ \sum_{\mathcal P} C_{\mathcal P \bar{\mathcal P}} \Gamma_{\mathcal P \bar{\mathcal P}} \right]\,,
\eeq
where $\sqrt s$ is the LHC $pp$ collision energy and $C_{\mathcal P \bar{\mathcal P}}$ parametrize the relevant parton luminosities. 

For illustration purposes in the following we consider in turn either $gg$ and $\gamma\gamma$ induced processes or alternatively $b\overline{b}$ and $\gamma\gamma$ rates at a benchmark mass of $M_S=750$~GeV. The remaining possibilities lie in between these two limiting cases considering the values of relevant parton luminosities (their values at
$\sqrt s=8$ TeV and 13 TeV LHC can be found e.g.~in~\cite{Franceschini:2015kwy}). 
In the former case given a 13 TeV cross-section $\sigma_{\gamma\gamma} \equiv \sigma (pp \to S ) \mathcal B_{\gamma\gamma}$
one obtains the relation
\beq
\label{xsecgg}
\frac{\Gamma_{\gamma \gamma} }{M_S} \frac{\Gamma_{gg} }{M_S} \simeq  1.4 \times 10^{-7} \frac{\sigma_{\gamma\gamma}}{\rm fb}\frac{\Gamma_S}{M_S} \, , 
\eeq
while for the latter we obtain 
\beq
\label{xsecbbbar}
\frac{\Gamma_{\gamma \gamma} }{M_S} \frac{\Gamma_{b\overline{b}} }{M_S}  \simeq 2.3 \times 10^{-5}  \frac{\sigma_{\gamma\gamma}}{\rm fb} \frac{\Gamma_S}{M_S} \, .
\eeq 
These relations define the phenomenological benchmarks for the resonance partial widths into gauge boson and quark final states, to be subjected to constraints from perturbative unitarity.

To make contact with the EFT unitarity discussion of the preceding subsections we use \eqs{Ggammagamma}{Ggg} and trade 
$\Lambda_ g$, $\Lambda_W$ and $\Lambda_B$ 
for $\Gamma_{gg}$, $\Gamma_{\gamma \gamma}$ and the ratio $r \equiv \Lambda_B/ \Lambda_W$. 
In particular, we get 
\begin{align}
\frac{1}{\Lambda^2_g} &= \frac{\Gamma_{gg}}{8 \pi \alpha^2_s M_S^3} \, , \\
\frac{1}{\Lambda^2_W} &= \frac{\Gamma_{\gamma \gamma}}{\pi \alpha_{\rm EM}^2 M_S^3} \left( \frac{r}{1+r}\right)^2 \, , \\
\frac{1}{\Lambda^2_B} &= \frac{\Gamma_{\gamma \gamma}}{\pi \alpha_{\rm EM}^2 M_S^3} \left( \frac{1}{1+r}\right)^2 \, ,
\end{align}
which inserted back into \eq{boundVV} yield 
\begin{equation}
\label{boundVV2}
\sqrt{s} \lesssim M_S \left( \frac{\Gamma_{gg}}{M_S} + f(r) \frac{\Gamma_{\gamma \gamma}}{M_S} \right)^{-1/2} \, ,
\end{equation}
with 
\beq
f(r) = \frac{ 3 r^2 s^{-4}_W + c_W^{-4}}{ (1+r)^2} \, .
\eeq 
Barring the fine-tuned region around $r=-1$ (corresponding to $1/\Lambda_\gamma = 0$), the function 
$f(r)$ has the global minimum 1.6 for $r=0.030$ and reaches asymptotically the maximum 57 for $r \to \pm \infty$.
%
Hence, we can set the following unitarity bounds
\begin{align}
\label{boundEFTgg}
\sqrt{s} &\lesssim  32 \, M_S  \left(\frac{\Gamma_{gg}/M_S}{10^{-3}}\right)^{-1/2} \, , \\
\label{boundEFTgammagamma}
\sqrt{s} &\lesssim  (13 \div 79) \, M_S \left(\frac{\Gamma_{\gamma \gamma}/M_S}{10^{-4}}\right)^{-1/2} \, ,
\end{align}
where the values 13 and 79 in the last equation correspond respectively to the boundary values $r \to \pm \infty$ and $0.030$.

Generally, these bounds can be interpreted as the indication of the mass scale of new degrees of freedom UV completing the effective low-energy description and regularizing (unitarizing) the 
amplitude growth. If $S$ is a member of a new strongly coupled sector (i.e.~a composite state)~\cite{Franceschini:2015kwy, Harigaya:2015ezk,Nakai:2015ptz,Pilaftsis:2015ycr,Belyaev:2015hgo,Bian:2015kjt,Molinaro:2015cwg,Barrie:2016ntq,Craig:2015lra,Draper:2016fsr,Redi:2016kip,Kamenik:2016izk}, the above results imply upper bounds on its compositeness scale.\footnote{It is an interesting question whether there could be an UV model where new dynamics shows up only at the scale of the ultimate 
unitarity violation, as e.g.~in \eq{boundEFTgammagamma}. A possibility would be for instance an SU$(N_{\rm TC})$ model of vector-like confinement (along the lines of Ref.~\cite{Redi:2016kip}) 
with a large $N_{\rm TC}$. Since the anomaly coefficients are enhanced by $N_{\rm TC}$, this would allow to obtain a parametrically large di-boson signal while keeping a relatively 
high confinenment scale $\Lambda_{\rm TC}$. A detailed study of the feasibility of such scenario goes beyond the scope of the present paper.} 
Unfortunately, in this context unless a prospective $\mathcal O(\rm TeV)$ mass di-boson resonance would have a very large di-boson decay width, the bounds do not appear strong enough to guarantee observable effects at LHC energies and a prospective future 50-100~TeV hadron-hadron collider~\cite{Assadi:2014nea, Tang:2015qga} would be called for.  On the other hand, in perturbative weakly-coupled realizations discussed in the next section, where $S$ remains an elementary particle in the UV, its couplings to SM gauge field strengths cannot be generated at the tree level. Thus one expects new dynamics to appear much below the above conservative unitarity estimates. 

In the case of quark scattering, we use \eq{Gbbbar} and \eq{matchEFT}. Thus the bound in \eq{boundbbbarEFT} translates into
\beq
\label{boundEFTbbbar}
\sqrt{s} \lesssim 2 \sqrt{3 \pi} v \left( \frac{\Gamma_{q\bar{q}}}{M_S} \right)^{-{1}/{2}}  
\simeq 4.8 \ \text{TeV} \left( \frac{\Gamma_{q\bar{q}}/M_S}{0.1} \right)^{-{1}/{2}}   \, , 
\eeq
where on the r.h.s. we have normalized the partial width in $q\bar q$ to a broad resonance scenario. 
Contrary to $S$ couplings to SM gauge field strengths, its couplings to SM fermions can be easily realized in weakly-coupled renormalizable models already at the tree level. In particular, this requires (a) $S$ mixing with the SM Higgs doublet, (b) embedding $S$ into an EW doublet with the quantum numbers of the SM Higgs, or (c) the introduction of new massive fermions mixing with the SM quarks and/or leptons. Case (a) is constrained by Higgs coupling measurements~\cite{Falkowski:2015swt, ATLASCONF2015044}.  In both remaining cases, the above result can be interpreted as an upper bound on the mass scale of the extra EW (and color) charged states present in the UV completions. Unfortunately, unless $S$ decay channels to SM quarks induce a sizable width, LHC energies will not necessarily be sufficient to exhaust these possibilities directly within the EFT. One should thus consider explicit UV realizations. In the case (b) which goes beyond the scope of this paper, precision Higgs boson and EW measurements can be used to provide additional handles~\cite{Angelescu:2015uiz, Becirevic:2015fmu, Han:2015qqj,Moretti:2015pbj,Han:2016bvl,Kamenik:2016tuv}. {Case (c) on the other hand, is covered in the next section.}

\begin{figure}[!ht]
  \begin{center}
    \includegraphics[width=.49\textwidth]{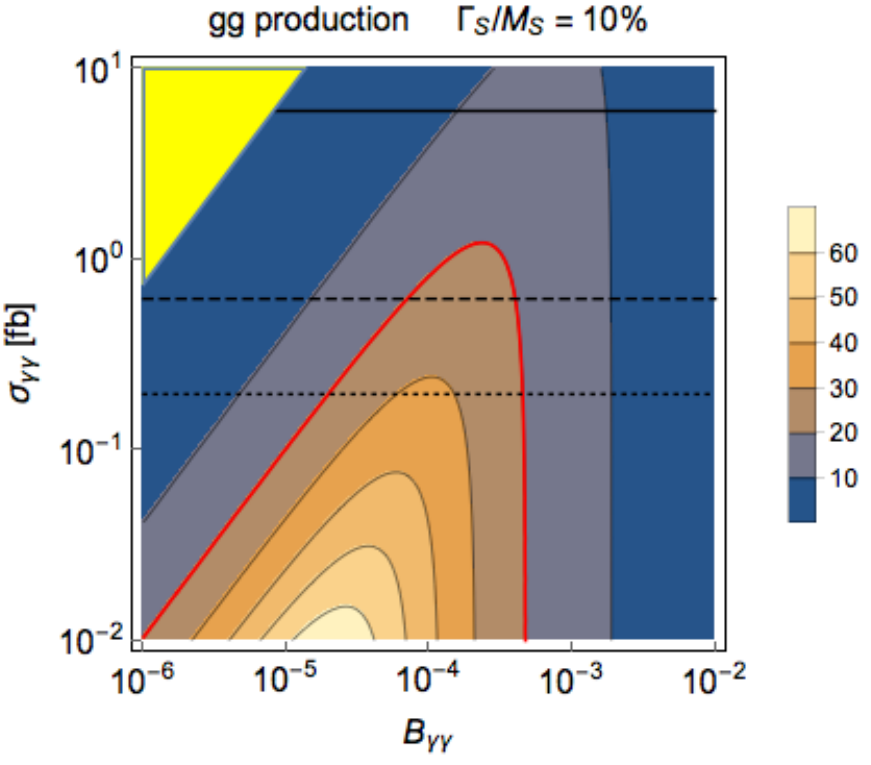}~~~~
    \includegraphics[width=.49\textwidth]{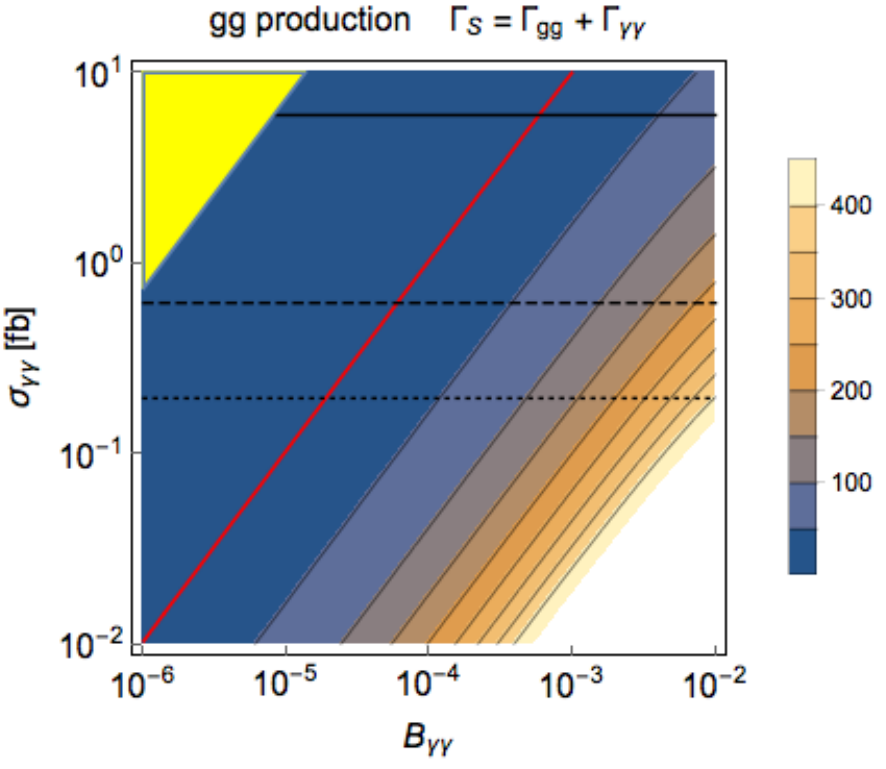}
   \\
   \vspace*{0.5cm}
    \includegraphics[width=.49\textwidth]{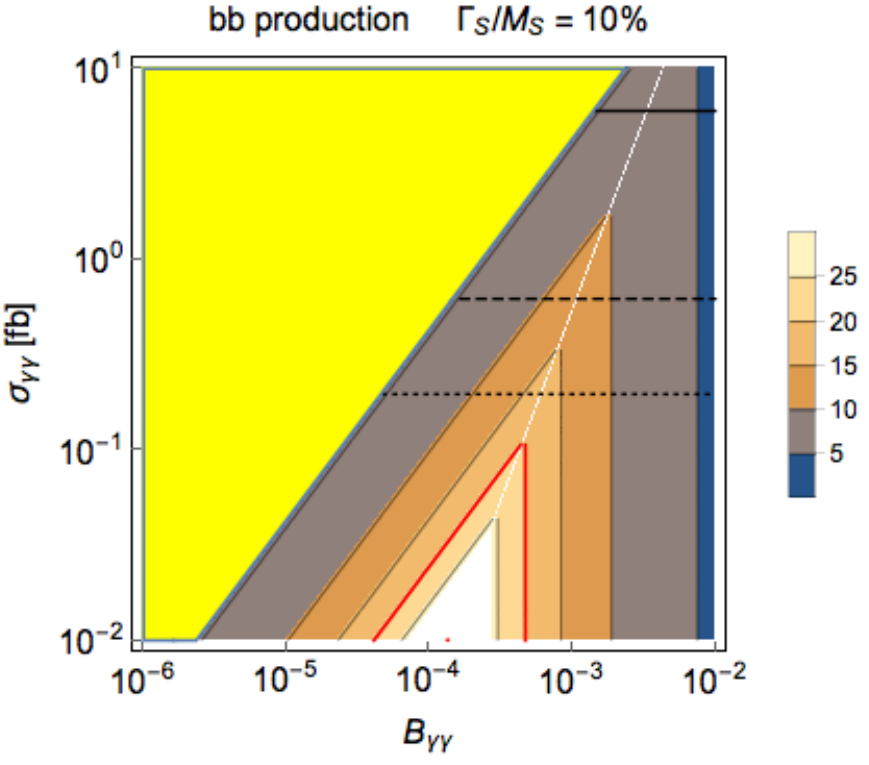}~~~~
    \includegraphics[width=.49\textwidth]{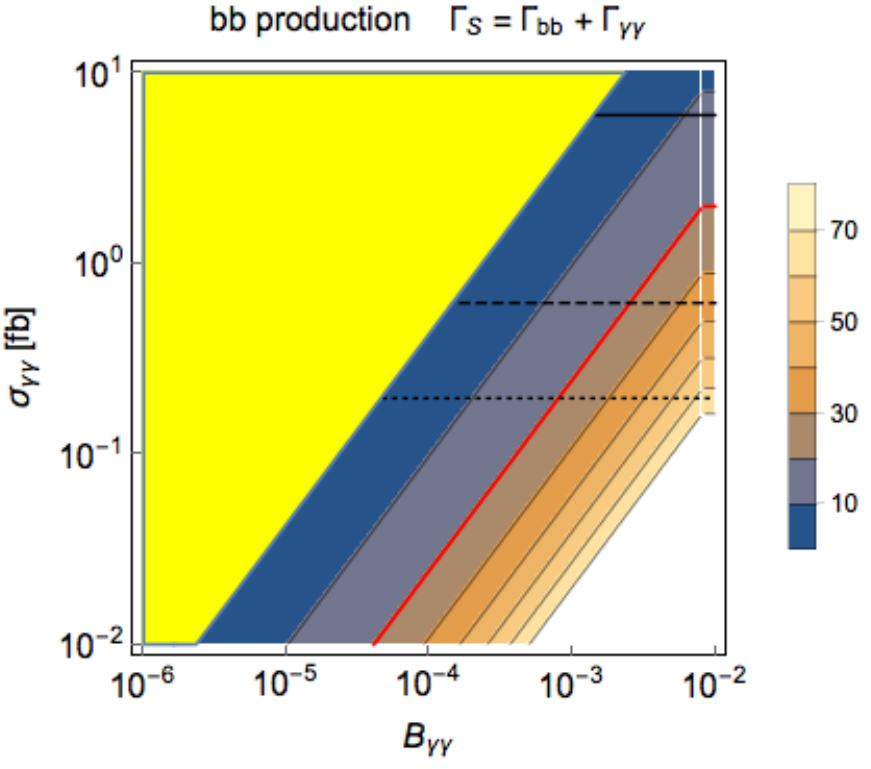}
  \end{center}
  \caption{\label{UBplot}
  Scale of unitarity violation $\Lambda_U$ in TeV in the $(\mathcal B_{\gamma\gamma}, \sigma_{\gamma\gamma})$ plane 
  (cf.~\eqs{LambdaUgg}{LambdaUbb}).  
  Upper/lower plots corresponding to $gg$/$b\overline{b}$ production, while 
  left/right plots to the large/small width scenario. As reference values we assume $M_S = 750$ GeV and $f(r)=30$. 
  The red curve denotes the new physics scale accessible at a futuristic 100 TeV collider, $\Lambda = 20$ TeV, 
  while the three horizontal lines from top to bottom are three reference cross-sections, namely $6$, $0.6$ and $0.2$ fb. 
  The yellow triangle on the top-left of each figure is the region in parameter space where $\Gamma_S / M_S > 10 \ \%$.}
\end{figure}

In \fig{UBplot} we display the scale of unitarity violation $\Lambda_U$ [TeV] in the 
$\mathcal B_{\gamma\gamma}$ vs.~$\sigma_{\gamma\gamma}$ plane, 
for either $gg$ or $b\overline{b}$ production and assuming either a broad or narrow resonance. 
In particular, for $gg$ production we have 
\beq 
\label{LambdaUgg}
\Lambda_U = M_S \left[ \frac{\Gamma_{gg}}{M_S} + f(r) \frac{\Gamma_{\gamma \gamma}}{M_S} \right]^{-1/2} \, ,
\eeq
while for $b\overline{b}$ production 
\beq 
\label{LambdaUbb}
\Lambda_U = \text{min} \left \{ 2 \sqrt{3 \pi} v \left( \frac{\Gamma_{b\overline{b}}}{M_S} \right)^{-1/2} , M_S 
\left[ f(r) \frac{\Gamma_{\gamma \gamma}}{M_S} \right]^{-1/2} \right \} \, . 
\eeq
As reference values we take $M_S = 750$ GeV and $f(r) = 30$. 
The horizontal lines from top to bottom indicate a cross-section signal of $6$, $0.6$ and $0.2$ fb, 
assuming the same significance of the signal over the three integrated luminosities $\int \mathcal{L} = 3.2$, $300$ and $3000$ fb$^{-1}$. 
The red curve denotes instead the reference value $\Lambda = 20$ TeV, corresponding 
to the typical squark-gluino reach of a futuristic 100 TeV collider \cite{Golling:2016gvc}, 
which applies in the case of coloured new physics generating the effective operators. 
Hence, if a signal is observed above the red curve it basically means that a 100 TeV collider could potentially probe 
the physics responsible for the restoration of unitarity.  We observe that such low-scale violation of unitarity are more readily obtained in the large width scenario and that for any given $\sigma_{\gamma\gamma}$ and $\mathcal B_{\gamma\gamma}$,  unitarity violation sets in earlier for $b\bar b$ induced production, compared to gluon fusion processes, due to much smaller PDFs.

\section{Weakly-coupled models}
\label{wcmodels}

In this section we consider explicit UV completions of the effective operators of \sect{reviewPWU}, 
capturing the main features of several proposed NP models, which have recently appeared in the literature. 
In particular, we will assume either fermion or scalar mediators\footnote{The case of vector mediators has been suggested and analyzed in Ref.~\cite{deBlas:2015hlv} within a simplified model. A complete renormalizable UV realization of this idea requires a non-trivial extension of the SM gauge sector, subject to many additional theoretical and experimental constraints. For this reason we do not consider such a possibility in our analysis.} and CP-even couplings 
(the CP-odd case leads to similar conclusions as far as concerns unitarity bounds). Moreover, 
we restrict ourselves to the cases of $b\bar b$, $gg$ and/or $\gamma\gamma$ decays and postulate different sets of fields which separately contribute to the relevant partial widths. Note that as far as perturbativity is concerned, this latter hypothesis 
leads to conservative bounds. Colored mediators are experimentally much more constrained, and their masses generally need to lie close to or above the TeV scale. On the other hand, much lighter uncolored mediators are still allowed, potentially leading to resonantly enhanced one-loop contributions to radiative $S$ 
decays~\cite{Bharucha:2016jyr,DiChiara:2016dez}. 

The first model comprises new fermionic mediators (see e.g.~\cite{Goertz:2015nkp}), all singlets under $SU(2)_L$. 
To this end, we introduce $N_Q$ copies of electromagnetic (EM) neutral vector-like QCD triplets $Q_A\sim(3,1,0)$ (with $A=1,\ldots,N_Q$) 
as well as $N_E$ copies of colorless vector-like fermions $E_B$ (with $B=1,\ldots,N_E$), 
with (hyper)charge $Y$ ($E_B\sim (1,1,Y)$). We assume the theory to be invariant under a 
$U(N_Q) \otimes U(N_E)$ global symmetry and the di-boson resonance is represented by a real scalar field $S$. 
The Lagrangian featuring the new fermions reads 
\begin{align} 
\label{LNF}
\mathcal{L}^{\rm{NF}} &= \overline{Q}_A i \slashed{D} Q_A + \overline{E}_B i \slashed{D} E_B \nonumber \\
&- \left( m_Q \overline{Q}_A Q_A + m_E \overline{E}_B E_B + y_Q S \overline{Q}_A Q_A + y_E S \overline{E}_B E_B \right) - V(S) \, ,
\end{align}
where the details of the scalar potential are not needed for our discussion. 

The second model we are going to consider involves instead new scalar mediators. 
In analogy to the previous case, we introduce 
$N_{\tilde{Q}}$ copies of EM neutral QCD scalar triplets $\tilde{Q}_A\sim(3,1,0)$ and
$N_{\tilde{E}}$ copies of colorless charged scalars $\tilde{E}_B\sim (1,1,Y)$, again all singlets under $SU(2)_L$. We also assume the theory to be invariant under a 
$U(N_{\tilde{Q}}) \otimes U(N_{\tilde{E}})$ global symmetry and the di-boson resonance is represented by a real scalar field $S$. 
The Lagrangian featuring the new scalars reads 
\begin{align} 
\label{LNS}
\mathcal{L}^{\rm{NS}} &= |D_\mu \tilde{Q}_A|^2 + |D_\mu \tilde{E}_B|^2 \nonumber \\ 
&- \left( m_{\tilde{Q}} \tilde{Q}^*_A \tilde{Q}_A + m_{\tilde{E}} \tilde{E}^*_B \tilde{E}_B 
+ A_Q S \tilde{Q}^*_A \tilde{Q}_A + A_E S \tilde{E}^*_B \tilde{E}_B \right) + \ldots \, ,
\end{align} 
where the ellipses stand for additional terms in the scalar potential which are irrelevant for our discussion. 

Focusing on the CP-even couplings, 
the contributions to $\Gamma_{\gamma\gamma}$ and $\Gamma_{gg}$ 
can now be written as~\cite{Franceschini:2015kwy}
\begin{align}
\label{PRgammagamma}
\frac{\Gamma_{\gamma\gamma}}{M_S} &= \frac{\alpha_{\rm EM}^2}{16\pi^3} 
\left| N_E Q^2_E y_E \sqrt{\tau_E} \mathcal{S}(\tau_E) + N_{\tilde{E}} Q^2_{\tilde{E}} \frac{A_E}{2 M_S} \mathcal{F}(\tau_{\tilde{E}})
\right|^2 \, , \\
\label{PRgg}
\frac{\Gamma_{gg}}{M_S} &= \frac{\alpha_s^2}{2\pi^3} 
\left| N_Q I_Q y_Q \sqrt{\tau_Q} \mathcal{S}(\tau_Q) + N_{\tilde{Q}} I_{\tilde{Q}} \frac{A_Q}{2 M_S} \mathcal{F}(\tau_{\tilde{Q}})
\right|^2  \, , 
\end{align}
where $\tau_i = 4 m^2_i/M^2_S$ (for $i = E,\tilde{E},Q,\tilde{Q}$), $I_{Q}=I_{\tilde{Q}}=1/2$ is the index of the QCD representation, 
while $Q_{E(\tilde{E})}$ is the EM charge of $E(\tilde{E})$. The loop functions read 
\begin{align}
\mathcal{S}(\tau) &= 1 + (1-\tau) \arctan^2(1/\sqrt{\tau-1}) \, , \\
\mathcal{F}(\tau) &= \tau \arctan^2(1/\sqrt{\tau-1}) -1 \, . 
\end{align}
In particular, in the limit of heavy particles $(\tau \to \infty)$, they decouple 
as $\mathcal{S}(\tau) \simeq 2/(3 \tau)$ and $\mathcal{F}(\tau) \simeq 1/(3 \tau)$.   
As a reference value we fix $M_S = 750$ GeV, $\alpha_s (M_S/2) = 0.1$, $\alpha_{\rm EM} = 1/137$ 
and set the masses of the mediators close to the current experimental bounds from direct searches,\footnote{Stable 
charged leptons must be heavier than about 400 GeV 
in order to avoid excessive Drell-Yan production \cite{Chatrchyan:2013oca,DiLuzio:2015oha}, 
while the bounds on long-lived colored particles are more model dependent 
due to non-perturbative QCD uncertainties 
and typically range from few hundreds of GeV to 1 TeV \cite{Aad:2013gva,Khachatryan:2015jha}.}
$m_{E,\tilde{E}} = 400$ GeV and $m_{Q,\tilde{Q}} = 1$ TeV, thus getting 
\begin{align}
\label{GammaNF}
\frac{\Gamma^{\rm{NF}}_{\gamma\gamma}}{M_S} &= 
7.8 \times 10^{-8} \ N_E^2 Q^4_E y_E^2  \, , \qquad\qquad\qquad\quad
\frac{\Gamma^{\rm{NF}}_{gg}}{M_S} =
2.7 \times 10^{-6} \ N_Q^2 y_Q^2 \\ 
\label{GammaNS}
\frac{\Gamma^{\rm{NS}}_{\gamma\gamma}}{M_S} &= 
1.2 \times 10^{-8} \ N_{\tilde{E}}^2 Q^4_{\tilde{E}} \left(\frac{A_E}{750 \ \rm{GeV}}\right)^2 \, , \qquad
\frac{\Gamma^{\rm{NS}}_{gg}}{M_S} =
2.6 \times 10^{-8} \ N_{\tilde{Q}}^2 \left(\frac{A_Q}{750 \ \rm{GeV}}\right)^2 \, , 
\end{align}
where we have separately considered the cases of new fermions and scalars.
For heavier mediator masses the rates decouple as powers of $1/\tau_i = M^2_S/(4 m^2_i)$ 
and thus even larger couplings are required. For this reason, 
 perturbativity bounds extracted using \eqs{GammaNF}{GammaNS} are understood to be conservative.  

Finally, we also consider a special case of the fermionic model, where at least one colored fermionic mediator has the SM gauge quantum numbers of the down-like right-handed SM quarks $\mathcal B \sim (3,1,-1/3)$ and mixes with the $b$-quark, in turn inducing $S\bar b b$ interactions.\footnote{Analogous cases for vector-like fermions mixing with other quark flavors can easily be derived using the results of~\cite{Fajfer:2013wca}.} The relevant $b-\mathcal B$ mixing Lagrangian is
\begin{align}
\label{VLmixing}
\mathcal L^{\mathcal B-b} &= \bar Q_3 i \slashed D Q_3 + \bar b_R i \slashed D b_R + \bar{\mathcal B} i \slashed D \mathcal B - (M_{\mathcal B} + \tilde y_{\mathcal B} S) \bar{\mathcal B} \mathcal B \nonumber\\
&- y_b \bar Q_3 H b_R - y_{\mathcal B} \bar Q_3 H \mathcal B_R - \tilde y_{b} \bar{ \mathcal B}_L S b_R + \rm h.c.\,,
\end{align}
where $Q_3= (t_L, b_L)$, we have used reparametrization invariance to rotate away a possible $\bar {\mathcal B} b_R$ mass-mixing term, and have also neglected small CKM induced mixing terms with the first two SM generations. 
In the following we assume all couplings to be real in accordance with the CP-even nature of $S$.
After EW symmetry breaking, the physical eigenstates $\mathcal B'$ and $b'$ are then given in terms of the above weak eigenstates as 
\beq
\left( \begin{array}{c} b_{L,R}' \\ \mathcal B_{L,R}' \end{array}\right) = \left( \begin{array}{cc} \cos\theta^{L,R}_{\mathcal B b} &  \sin\theta^{L,R}_{\mathcal B b} \\ -\sin\theta^{L,R}_{\mathcal B b} & \cos\theta^{L,R}_{\mathcal B b} \end{array} \right) \left( \begin{array}{c} b_{L,R} \\ \mathcal B_{L,R} \end{array}\right)\,,
\eeq
where
\begin{align}
\tan 2 \theta^L_{\mathcal B b} &= \frac{\sqrt 2 v y_{\mathcal B } M_{\mathcal B}}{M_{\mathcal B}^2 - \left[ y_b^2 + y_{\mathcal B }^2 \right] v^2/2}\,, \\
\tan 2 \theta^R_{\mathcal B b} &= \frac{v^2 y_{b} y_{\mathcal B } }{M_{\mathcal B}^2 - \left[ y_b^2 - y_{\mathcal B}^2 \right] v^2/2}\,, 
\end{align}
and the masses are related via
\beq
m_b m_{\mathcal B} = M_{\mathcal B} y_b \frac{v}{\sqrt 2} \,, \qquad m_b^2 + m_{\mathcal B}^2 = M_{\mathcal B}^2 + \frac{v^2}{2} \left[  y_b^2 + y_{\mathcal B b}^2 \right]\,.
\eeq
In this basis, the $S$ interactions with $b'$ and $\mathcal B'$ are
\begin{align}
- \mathcal L^{\mathcal B - b} &\ni  S \left[  \bar {\mathcal B'} \mathcal B' \cos\theta^L_{\mathcal B b} ( \cos\theta^R_{\mathcal B b}  \tilde y_{\mathcal B} -  \sin \theta^R_{\mathcal B b}   \tilde y_b )  + \bar b' b' \sin\theta^L_{\mathcal B b}( \sin\theta^R_{\mathcal B b}  \tilde y_{\mathcal B} + \cos \theta^R_{\mathcal B b}  \tilde y_b)\right. \nonumber\\
&\left. + \bar {\mathcal B'_R} b'_L  \sin \theta^L_{\mathcal B b}  (\cos \theta^R_{\mathcal B b}  \tilde y_{\mathcal B} -  \sin\theta^R_{\mathcal B b}  \tilde y_b  )+ \bar {\mathcal B'_L} b'_R  \cos \theta^L_{\mathcal B b}   (\sin \theta^R_{\mathcal B b}  \tilde y_{\mathcal B} + \cos\theta^R_{\mathcal B b}  \tilde y_b   ) + {\rm h.c.} \right]  \,.
\end{align}
The $\theta^L_{\mathcal B b}$ mixing angle is constrained by EW precision measurements to $\sin\theta^L_{\mathcal B b} = 0.05(4) $~\cite{Fajfer:2013wca}, while $\theta^R_{\mathcal B b}$ is parametrically further suppressed as $\theta^R_{\mathcal B b}  \sim (m_b / m_{\mathcal B}) \theta^L_{\mathcal B b}$. The $S\to b \bar b$ decay width can thus be written compactly as
\beq
\label{Sbb}
\frac{\Gamma_{b\bar b}}{M_S}  = \frac{3}{8\pi}  \sin^2 \theta^L_{\mathcal B b}     \tilde y_b^2 = 3\times 10^{-4} \left(\frac{\sin \theta^L_{\mathcal B b}}{0.05}\right)^2   \tilde y_b^2 \,,
\eeq
up to terms suppressed as $m_b^2/\left\{M_S^2, m^2_{\mathcal B}\right\}$\,. Note that contrary to the loop induced decay modes, $\Gamma_{b\bar b}$ does not explicitly depend on the mediator mass. On the other hand, its implicit dependence through $\theta_{\mathcal B b}^{L} \sim v/m_{\mathcal B}$ is well constrained experimentally. The resulting unitarity constraints based on \eq{Sbb} and saturating the upper bound on  $\theta_{\mathcal B b}^{L} $ can thus again be considered as conservative.

\subsection{Single fermion case}
\label{fermmed}

\begin{figure}[!ht]
  \begin{center}
    \includegraphics[width=.42\textwidth]{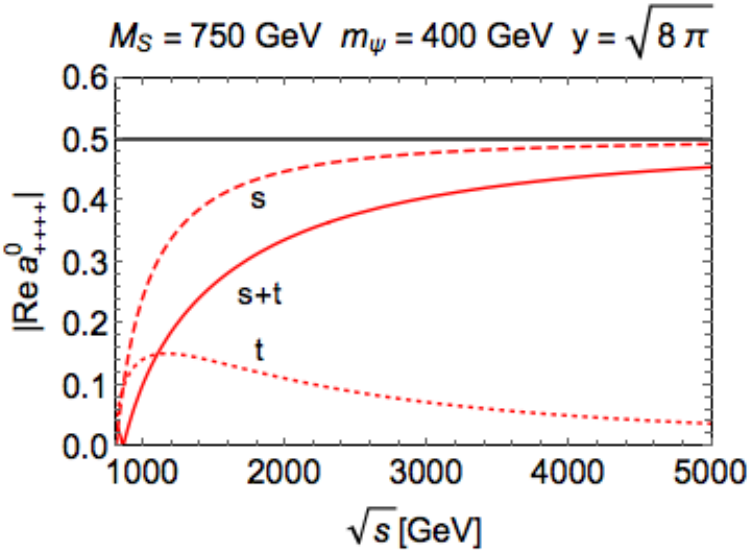}~~~~
    \includegraphics[width=.42\textwidth]{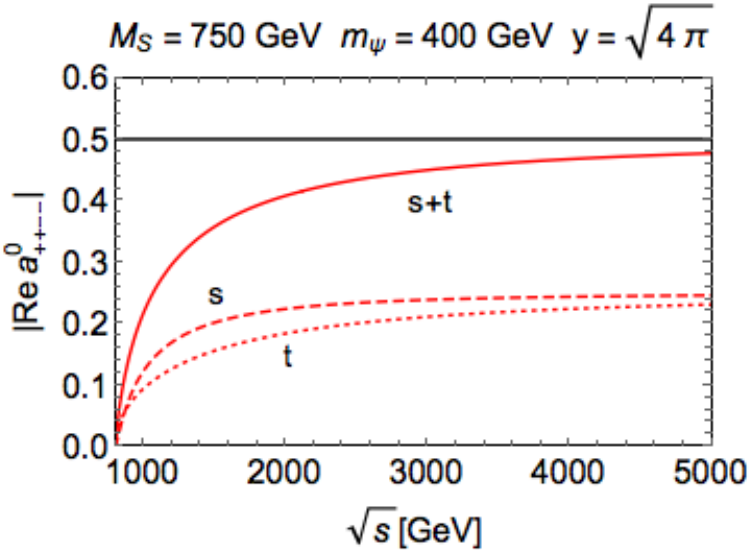}
  \end{center}
  \caption{\label{FermScat}
  Full kinematical dependence of $|\Re a^0_{++++}|$ (left panel) and $|\Re a^0_{++--}|$ (right panel), for the reference values 
  $M_S = 750$ GeV, $m_\psi = 400$ GeV and $y = \sqrt{8 \pi}$ (left panel) and $y = \sqrt{4 \pi}$ (right panel). 
  Dashed, dotted and full (red) lines represent respectively 
  $s$-, $t$-channel and full contribution to the partial wave. Asymptotically, for $\sqrt{s} \gg M_S, m_\psi$, 
  the values $|\Re a^0_{++++}| \simeq \frac{1}{2}$ and $|\Re a^0_{++--}| \simeq \frac{1}{2}$ are reached.}
\end{figure}

Let us first consider a simplified model featuring a real scalar singlet $S$ and a non-colored Dirac fermion $\psi$, 
with the interaction Lagrangian
\beq
\label{intSpsibarpsi}
\mathcal{L}_I \supset - y S \overline{\psi} \psi \, . 
\eeq 
We denote the masses of $S$ and $\psi$, respectively as $M_S$ and $m_\psi$. 
Focusing on the $J=0$ sector, the most relevant scattering amplitude is given by $\psi \overline{\psi} \to \psi \overline{\psi}$ (cf.~\app{psipsibarpsipsibar}).
In particular, the matrix of scattering amplitudes in the ($++,--$) helicity basis\footnote{$+-$ and $-+$ have zero projection on the $J=0$ sector.} is found to be
\beq
\label{TFSfull}
\mathcal{T} = -y^2
\left( 
\begin{array}{cc}
\frac{4 (p^3)^2}{s-M^2_S} + \frac{4 m^2 \cos^2\frac{\theta}{2}}{t - M^2_S} & \frac{4 (p^3)^2}{s-M^2_S} - \frac{4 E^2 \cos^2\frac{\theta}{2}}{t - M^2_S} \\
\frac{4 (p^3)^2}{s-M^2_S} - \frac{4 E^2 \cos^2\frac{\theta}{2}}{t - M^2_S} & \frac{4 (p^3)^2}{s-M^2_S} + \frac{4 m^2 \cos^2\frac{\theta}{2}}{t - M^2_S} 
\end{array}
\right) 
\overset{\sqrt{s} \ \gg \ M_S, \, m_\psi}{\simeq}
- y^2 
\left( 
\begin{array}{cc}
1 & 2 \\
2 & 1
\end{array}
\right) 
\, ,
\eeq
where in the last step we took the high-energy limit. 
The projection on the $J=0$ partial waves is readily obtained by applying \eq{PWproj}. 
We report here the expression in the high-energy limit 
(for the full expression see \eqs{a0++++full}{a0++--full} in \app{psipsibarpsipsibar})
\beq
a^0 \simeq 
- \frac{y^2}{16 \pi} 
\left( 
\begin{array}{cc}
1 & 2 \\
2 & 1
\end{array}
\right) 
\, . 
\eeq
The tree-level unitarity bound (cf.~\eq{UnitBound}) relative to the 
highest eigenvalue of the partial wave matrix yields
\beq 
\label{unityuk}
 y \lesssim \sqrt{\frac{8 \pi}{3}} \, .
\eeq
The behaviour of $|\Re a^0_{++++}|$ ($|\Re a^0_{++--}|$) with the full kinematical dependence 
is displayed in 
\fig{FermScat}, for the reference values $M_S = 750$ GeV, $m_\psi = 400$ GeV and 
$y = \sqrt{8 \pi}$ ($y = \sqrt{4 \pi}$).  

\begin{figure}[!ht]
  \begin{center}
    \includegraphics[width=.32\textwidth]{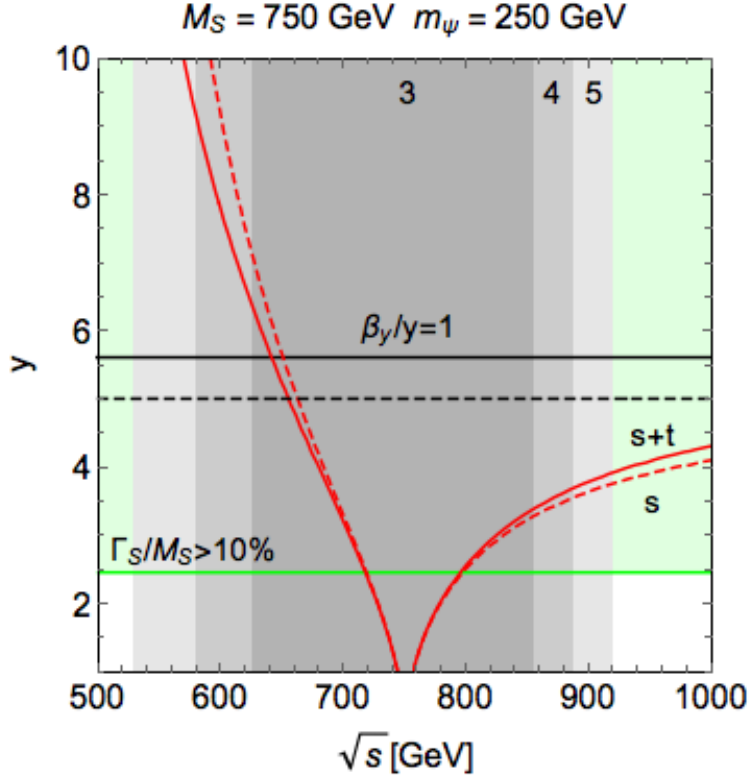}~~
    \includegraphics[width=.32\textwidth]{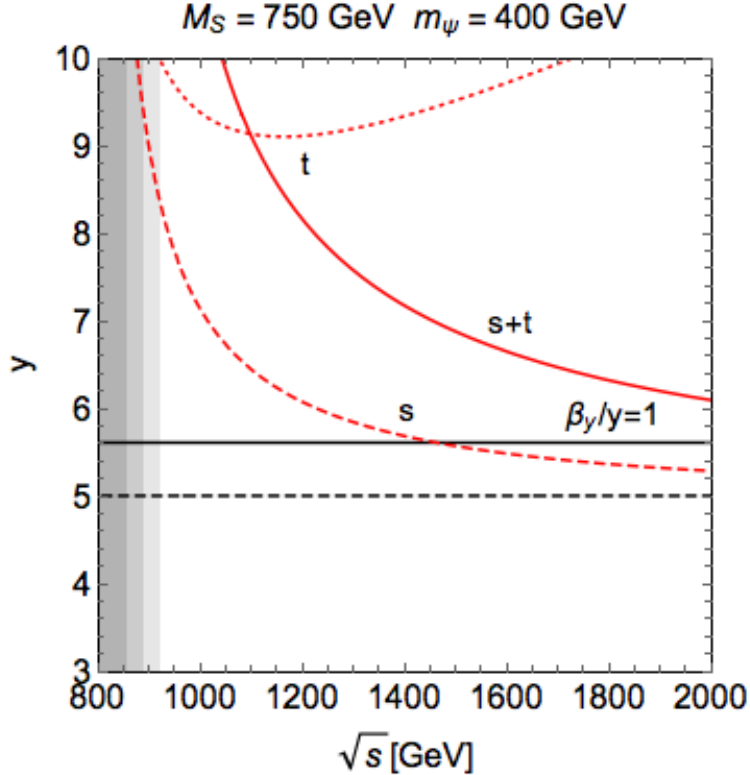}~~
    \includegraphics[width=.32\textwidth]{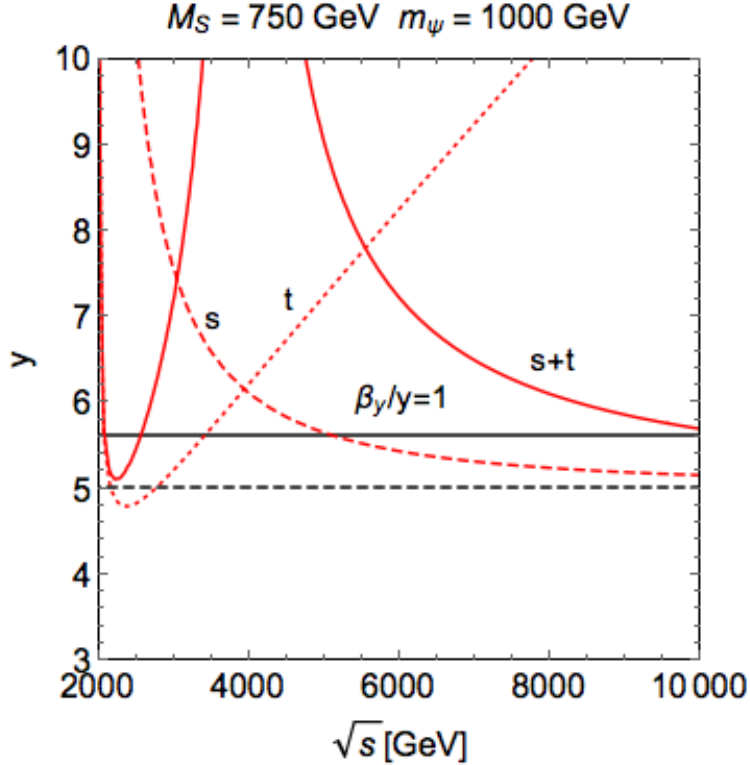}
  \end{center}
  \caption{\label{FS_yvssqrts_pppp}
  Saturation of the tree-level unitarity bound, $|\Re a^0_{++++}|=1/2$, in the $(\sqrt{s},y)$ plane 
  for $M_S = 750$ GeV and the three reference values $m_\psi = 250$, 400 and 1000 GeV. 
  Dashed, dotted and full (red) lines denote respectively the $s$-, $t$-channel 
  and full contribution to the partial wave. 
  The light-green shaded area in the first plot corresponds to the region where 
  $\Gamma_S/M_S > 10 \%$,  
  while the grey-level vertical bands are contours of possible finite width effects 
  defined in \eq{cutsa0} with $\alpha=3$, 4, 5.  
  The dashed (black) horizontal line indicates the asymptotic value $y = \sqrt{8 \pi} \simeq 5$, 
  while the full (black) line is the perturbativity bound obtained from the 
  RGE criterium $\beta_y / y < 1$ (cf.~\eq{RGEbound}). 
  }
\end{figure}

In \fig{FS_yvssqrts_pppp} we show 
the tree-level unitarity bound, restricted for simplicity to the ++++ helicity channel 
and for the three reference values $m_\psi = 250$, 400 and 1000 GeV. 

The above discussion prompts us to investigate resonance width effects, which can also become important very close to the scattering poles and effectively regulate the formally diverging tree-level amplitudes. Since such effects necessarily go beyond the tree-level approximation (they can be viewed as the absorptive part of the resummed self-energy contributions of $S$), we do not attempt to include 
them explicitly.\footnote{For a different approach see Refs.~\cite{Dawson:1988va,Dawson:1989up}.} 
Instead we superimpose contours of constant $s$ (in shades of grey) where the (on-shell) width effects 
parametrized as\footnote{For a similar approach see 
Refs.~\cite{SchuesslerDiplom,Schuessler:2007av,Betre:2014fva}.} 
\beq
\label{cutsa0}
\alpha = \frac{ |s-M_S^2|}{\Gamma_S M_S}\,,
\eeq
are expected to become important. Unitarity constraints derived in such regions cannot be considered meaningful.
The parameter $\alpha$ in \eq{cutsa0} can be viewed as 
a measure of the relative error $\Delta$ introduced by using the tree-level propagator in the squared amplitude 
instead of  one corrected in a Breit-Wigner approximation. In particular, we have $\alpha = \sqrt{1/\Delta-1}$. 
So, for example, $\alpha = 3$ corresponds to $\Delta = 10\%$. 
For concreteness we fix $\Gamma_S/M_S=0.10$. Note that due to the scaling of \eq{cutsa0}, smaller $S$ decay widths can only lead to more stringent constraints (derived closer to the resonance poles). The bounds derived in this way can thus be considered conservative.

For $m_\psi = 250$ GeV, $S$ can directly decay into $\psi \overline{\psi}$, thus giving the 
following contribution to the total decay rate 
\beq 
\label{GammaSfermion}
\Gamma_S = \frac{y^2}{8\pi} M_S \left( 1 - \frac{4 m^2_\psi}{M_S} \right)^{3/2} \, .
\eeq
In fact the requirement $\Gamma_S/M_S < 10 \%$ is always more constraining than the tree-level unitarity bound 
whenever the $s$-pole resonance is above threshold, $M_S > 2 m_\psi$ (cf.~shaded light-green region 
in the first plot of \fig{FS_yvssqrts_pppp}). On the other hand, for cases where the 
$s$-pole resonance is below threshold, tree-level unitarity is violated well above 
the region where resonance width effects are relevant.


It is interesting to compare the tree-level unitarity bounds in \fig{FS_yvssqrts_pppp} 
with those obtained via the RGE criterium \cite{Goertz:2015nkp} 
\beq 
\label{RGEbound}
\frac{\beta_y}{y} = \frac{5  y^2}{16 \pi^2} < 1 \, . 
\eeq
The latter agrees up to an $\mathcal{O}(1)$ factor with the bound based on tree-level unitarity in the asymptotic high-energy regime $y < \sqrt{8 \pi}$. 

Finally we note that in addition to $\psi\bar \psi$ scattering, in bounding tree-level unitarity within the fermionic mediator model one can also consider other elastic channels, such as $\psi S$ or $\psi \psi$. It turns out however, that the corresponding $J=0$ partial wave amplitudes vanish in the $\sqrt{s} \to \infty$ limit and also do not receive possible enhancements due to nearby 
$s$-channel resonance poles, thus leading to no additional constraints.

\subsection{Single scalar case}
\label{scalarmed}

\begin{figure}[!ht]
  \begin{center}
    \includegraphics[width=.42\textwidth]{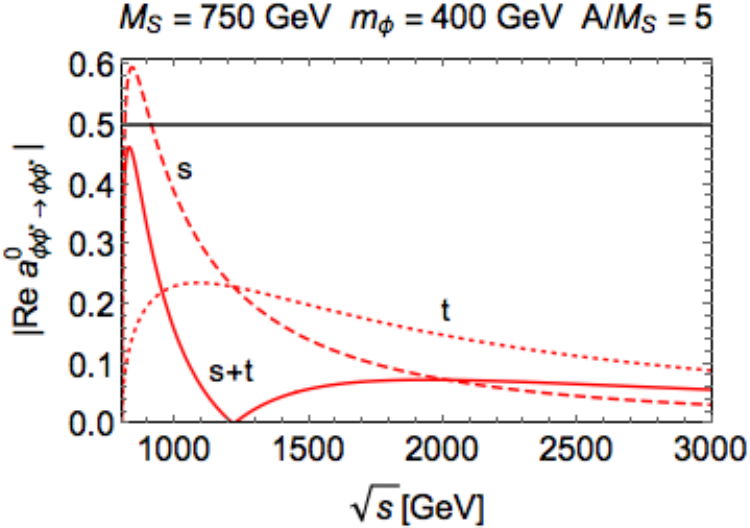}~~~~
    \includegraphics[width=.42\textwidth]{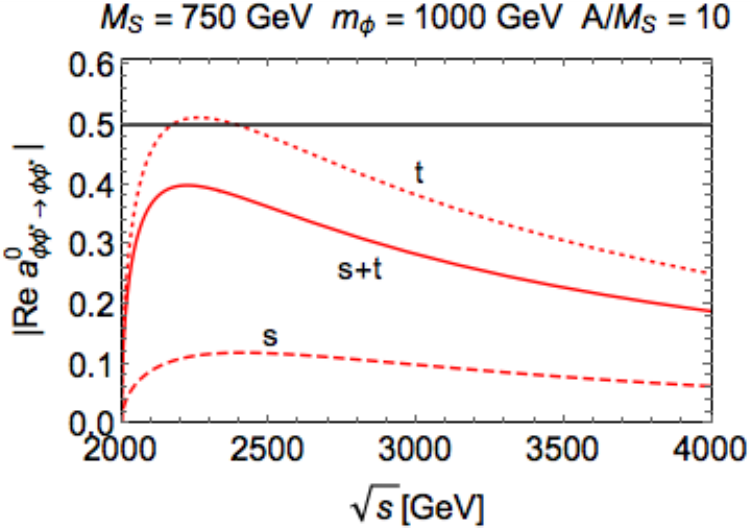} 
\end{center}
  \caption{\label{SS_avssqrts}
   Full kinematical dependence of $|\Re a^0_{\phi \phi^* \to \phi \phi^*}|$, for the reference values 
  $M_S = 750$ GeV, $m_\phi = 400$ GeV and $A/M_S = 5$ (left panel). 
  Same for $m_\phi = 1000$ GeV and $A/M_S = 10$ (right panel). 
  Dashed, dotted and full (red) lines represent respectively 
  $s$-, $t$-channel and the full contribution to the partial wave. Asymptotically, for $\sqrt{s} \gg M_S, m_\phi$, 
   $|\Re a^0_{\phi \phi^* \to \phi \phi^*}|$ approaches zero for any value of the coupling $A$.}
\end{figure}

Let us next consider the scalar resonance $S$ interacting with a complex scalar field $\phi$ via 
\beq
\label{ASpps}
\mathcal{L}_I \supset - A S \phi^* \phi \, ,
\eeq
where $A$ is a massive coupling and the masses of $S$ and $\phi$ are denoted as $M_S$ and $m_\phi$, 
respectively. 
The amplitude for the $\phi \phi^* \to \phi \phi^*$ scattering reads
\beq 
\mathcal{T}_{\phi \phi^* \to \phi \phi^*} = - A^2 \left( \frac{1}{s - M_S^2} + \frac{1}{t - M_S^2} \right) \, .
\eeq
Correspondingly, the $J=0$ partial wave is found to be 
\beq 
\label{a0scalar}
a^0_{\phi \phi^* \to \phi \phi^*} = - A^2 \frac{\sqrt{s (s - 4 m_\phi^2)}}{16 \pi s}
\left( \frac{1}{s - M_S^2} - \frac{\log\frac{s - 4 m_\phi^2 + M_S^2}{M_S^2}}{s - 4 m_\phi^2} \right) \, ,
\eeq
whose behaviour is shown in the left (right) panel of 
\fig{SS_avssqrts} for the reference values $M_S = 750$ GeV, $m_\phi = 400$ GeV 
($1000$ GeV) and 
$A/M_S = 5$ ($10$). 

Note that, differently from the fermion mediators' case, 
the unitarity bound is never relevant in the high-energy regime $\sqrt{s} \gg M_S, m_\phi$.
Such situation is expected since the scalar interaction in Eq.~\eqref{ASpps} is in the form of a relevant operator, whose tree-level contribution to $a^0$ vanishes as $1/s$ in the $s\to\infty$ limit.   
Thus tree-level unitarity in this case cannot be used to bound the validity of the leading order perturbative description at high energies. 
It can nonetheless identify problematic kinematical regions in vicinity of scattering poles.

\begin{figure}[!ht]
  \begin{center}
    \includegraphics[width=.32\textwidth]{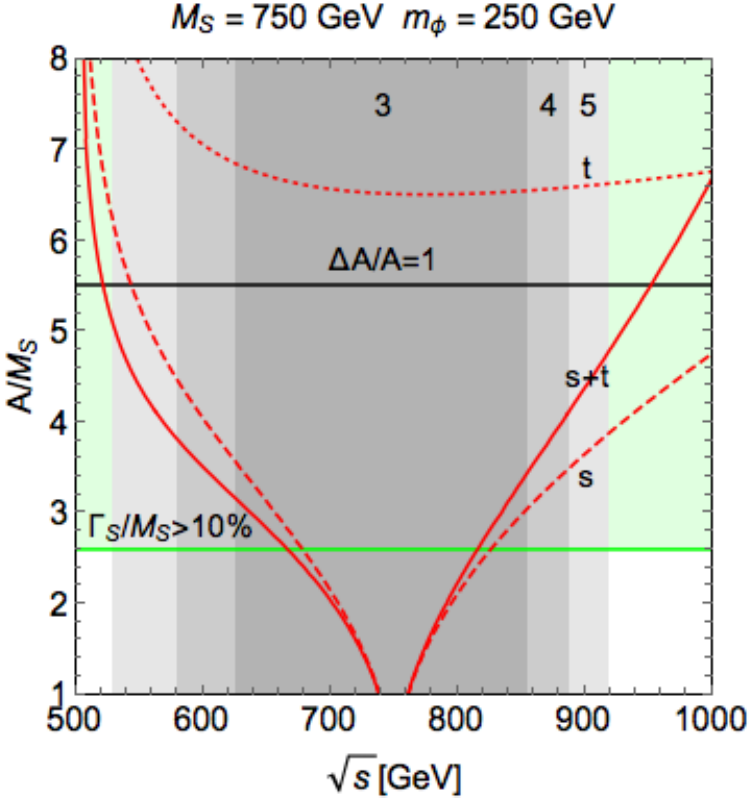}~~
    \includegraphics[width=.32\textwidth]{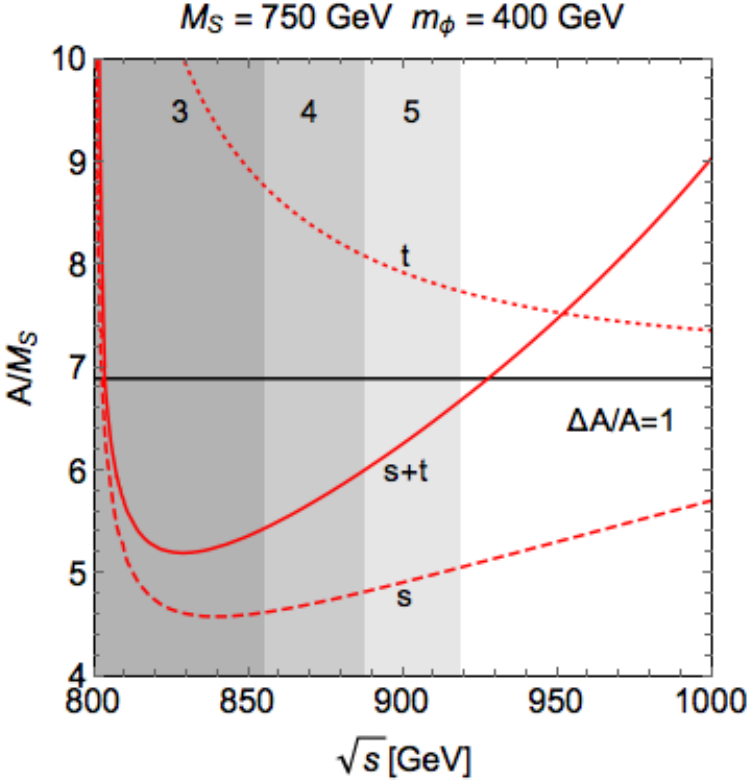}~~
    \includegraphics[width=.32\textwidth]{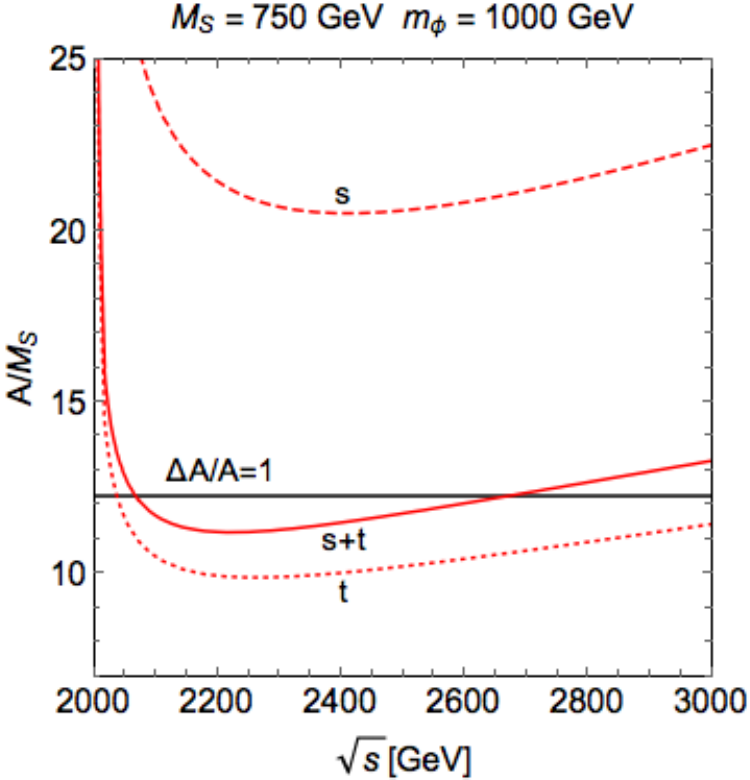}
  \end{center}
  \caption{\label{SS_AoMvssqrts}
  Saturation of the tree-level unitarity bound, $|\Re a^0_{\phi \phi^* \to \phi \phi^*}|=1/2$, 
  in the $(\sqrt{s},A/M_S)$ plane 
  for $M_S = 750$ GeV and the three reference values $m_\phi = 250$, 400 and 1000 GeV. 
  Dashed, dotted and full (red) lines represent respectively 
  $s$-, $t$-channel and the full contribution to the partial wave. 
  The light-green shaded area in the first plot corresponds to the region where 
  $\Gamma_S/M_S > 10 \%$, 
  while the grey-level vertical bands are the cuts due to finite width effects 
  defined in \eq{cutsa0} with $\alpha=3$, 4, 5.  
  The full (black) line is the perturbativity bound obtained from the 
  finite trilinear vertex correction $\Delta A / A < 1$ (cf.~\eq{pertboundfinite}). 
  }
\end{figure}

\fig{SS_AoMvssqrts} shows the unitarity bound 
for the three reference values $m_\phi = 250$, 400 and 1000 GeV.  
For $m_\phi = 250$ GeV, the $S \to \phi \phi^*$ decay channel contributes to the total width of $S$ via  
\beq 
\label{GammaSscalar}
\Gamma_S = \frac{1}{16 \pi} \frac{A^2}{M_S} \sqrt{1-\frac{4 m_\phi^2}{M_S^2}} \, .
\eeq
Analogously to the fermionic case, whenever the $s$-pole resonance is above threshold, $M_S > 2 m_\phi$,  
the requirement $\Gamma_S/M_S < 10 \%$ is always more constraining than the tree-level unitarity bound 
(cf.~light-green shaded area in the first plot of \fig{SS_AoMvssqrts}).  
Below threshold, the issue of the $s$-pole resonance width is treated in a similar way as for the fermionic case, by identifying and avoiding
 kinematical regions in $\sqrt{s}$  via \eq{cutsa0} where finite width effects can become important. 
For $m_{\phi}=400$ $(1000)$ GeV,  tree-level unitarity is then violated for values of $A/M_S \gtrsim 6.6$ ($11$), 
at scales of $\sqrt{s} \simeq 920$ GeV ($2.2$ TeV). 

Comparing the above tree-level unitarity bound with a complementary perturbativity criterium, we notice that in this case the RGEs cannot be used since, $A$ being associated to a relevant operator, 
by dimensional reasons it cannot enter its beta function alone. 
However, $A$ does give a finite perturbative correction to the 
trilinear scalar vertex $S\phi\phi^*$. By evaluating the one-loop correction at zero external momentum we find 
\beq 
\label{pertboundfinite}
\Delta A = \frac{1}{16 \pi^2} \frac{A^3}{m_\phi^2 - M^2_S} 
\left( 1+ \frac{M^2_S \log{\frac{M^2_S}{m_\phi^2}} }{m_\phi^2 - M^2_S} \right) \, .
\eeq
In the $m_\phi \gg M_S$ limit we have 
\beq 
\Delta A = \frac{1}{16 \pi^2}  \frac{A^3}{m^2_\phi} + \mathcal{O}\left(\frac{M_S}{m_\phi}\right)^2 \, ,
\eeq
while for $M_S \gg m_\phi$
\beq 
\Delta A = \frac{1}{16 \pi^2}  \frac{A^3}{M_S^2} \left ( 1 + \log{\frac{m_\phi^2}{M^2_S}} \right) + \mathcal{O}\left(\frac{m_\phi}{M_S}\right)^2 \, .
\eeq 
We can hence define a perturbativity criterium via the relation $\Delta A / A < 1$. 
In any of the two limits above, the bound $\Delta A / A < 1$ is approximately given 
by\footnote{A similar estimate of the onset of the non-pertubative regime, based on naive dimensional analysis, has been suggested in \cite{Baratella:2016daa}.}
\beq 
\label{approxDeltaoA}
\frac{A}{\text{max} \, \{ m_\phi, M_S \}} < 4 \pi \, , 
\eeq 
which agrees within an $\mathcal{O}(1)$ factor with the bound based on tree-level unitarity (cf.~also \fig{SS_AoMvssqrts}). 

We also note that a conceptually different bound could be inferred by requiring that $A$ does not destabilize 
too much the $d=2$ operators.\footnote{This is essentially a hierarchy problem, not related to perturbativity.} 
For instance, by inspecting the beta function of $M^2_S$ (see e.g.~\cite{Martin:1993zk})
\beq
\beta_{M^2_S} = \frac{A^2}{8 \pi^2} \, , 
\eeq
we might require $\beta_{M^2_S} / M^2_S = A^2 / 8 \pi^2 < 1$, which yields a bound very similar to that in \eq{approxDeltaoA}. 
On the other hand, an interesting feature of the mass-hierarchy bound is that, unlike the one obtained via 
the finite vertex correction, it gets enhanced by the multiplicity $N$ of fields $\phi$ coupling to $S$,  
via the replacement $A^2 \to N A^2$. 

Finally, in addition to the $\phi \phi^*$ channel, one could also consider the $\phi S$ or $\phi \phi$ scatterings. 
However, for reasons similar to the fermionic case, these processes do not lead to additional constraints and we do not discuss them any longer. 

\subsection{Generalization in flavor space}

The  results of the previous two subsections can be readily generalized to the case of $N$ copies of the mediators. 
The same conclusions apply for fermion and scalar mediators, but for definiteness we are going to 
explicitly discuss them for fermions only. 
To this end, let us consider $N$ copies of fermion fields, $\psi_i$ ($i=1,\ldots,N$), interacting via the Lagrangian term 
\beq
\mathcal{L}_I \supset - y_{ij} S \overline{\psi}_i \psi_j \, , 
\eeq 
where $y_{ij}$ is understood in the mass basis. 
Let us assume then some flavor structures for $y_{ij}$ and study the corresponding form of the unitarity bound: 
\begin{enumerate}
\item $y_{ij} = y$ ($\forall$ $i$ and $j$) 

In such a case the amplitude matrix in \eq{TFSfull} gets generalized into 
\beq 
\mathcal{T} \otimes J_N \, ,
\eeq
where $\otimes$ denotes Kronecker product and $J_N$ is the $N$-dimensional matrix made all by 1's. 
Since the only non-zero eigenvalue of $J_N$ is equal to $N$ (recall that $J_N$ is a rank-1 matrix), 
all the results of the 
previous section are readily generalized by the replacement $y \to \sqrt{N} y$.

\item $y_{ij} = y \delta_{ij}$

This case corresponds to the weakly-coupled models discussed at the beginning of \sect{wcmodels}.
The Lagrangian features an extra $U(N)$ global symmetry which can be conveniently 
used to label the irreducible sectors of the $\psi \overline{\psi} \to \psi \overline{\psi}$ scattering amplitudes. 
Since $N \otimes \bar{N} = \mathbf{1} \oplus \text{Adj}_N$, a general two-particle state $|\psi_i \overline{\psi}_j \rangle$ 
can be decomposed into a singlet channel
\beq 
|\psi \overline{\psi}\rangle_\mathbf{1} 
= \frac{1}{\sqrt{N}} \sum_i |\psi_i \overline{\psi}_i\rangle \, ,
\eeq
and an adjoint one
\beq 
|\psi \overline{\psi}\rangle_\mathbf{\text{Adj}}^A 
= T^A_{ij} |\psi_i \overline{\psi}_i\rangle \, ,
\eeq
where $T^A$, with $A = 1, \ldots, N^2-1$, are $SU(N)$ generators in the fundamental representation 
(in the normalization $\Tr T^A T^B = \delta^{AB}$)
and we properly took into account the normalization of the states.

Due to the specific flavor structure, $y_{ij} = y \delta_{ij}$, 
one has 
\beq 
\langle \psi_k \overline{\psi}_l |S|\psi_i \overline{\psi}_j\rangle =  i \mathcal{T}_s \, \delta_{ij} \delta_{kl} + i \mathcal{T}_t \, \delta_{ik} \delta_{jl} \, ,
\eeq
where $\mathcal{T}_s$ and $\mathcal{T}_t$ denote respectively the $s$- and $t$-channel 
contribution to the scattering amplitudes in \eq{TFSfull}. 

Let us hence discuss in turn the non-zero scattering amplitudes. 
For the singlet-singlet channel one finds
\beq 
\label{singletchamp}
_\mathbf{1} \langle \psi \overline{\psi}  |S|\psi \overline{\psi}\rangle_\mathbf{1} 
= \frac{1}{N} \sum_{ik} \langle \psi_k \overline{\psi}_k |S|\psi_i \overline{\psi}_i\rangle 
= \frac{1}{N} \sum_{ik} (i \mathcal{T}_s \, \delta_{ii} \delta_{kk} + i \mathcal{T}_t \, \delta_{ik} \delta_{ik})
= i \mathcal{T}_s \, N + i \mathcal{T}_t \, .
\eeq
In the asymptotic limit, $\sqrt{s} \gg M_S, m_\psi$, the $t$-channel decouples and one recovers the same 
multiplicity suppression in the unitarity bound, 
as in case 1.  
The results in the low-energy region are instead displayed in \fig{FSflav_yvssqrts_pppp}, 
which shows the tree-level unitarity bound in the $(\sqrt{s},\sqrt{N}y)$ plane, for different values of $N$. 
Notice that, in this normalization, the $s$-channel contribution is not affected by $N$, 
while the $t$-channel contribution is suppressed like $1/N$ (cf.~\eq{singletchamp}). 
Hence, for large enough $N$ the unitarity bound coincides with the $s$-channel one 
and becomes relevant only in the asymptotic region $\sqrt{s} \gg M_S, m_\psi$.

\begin{figure}[!ht]
  \begin{center}
    \includegraphics[width=.40\textwidth]{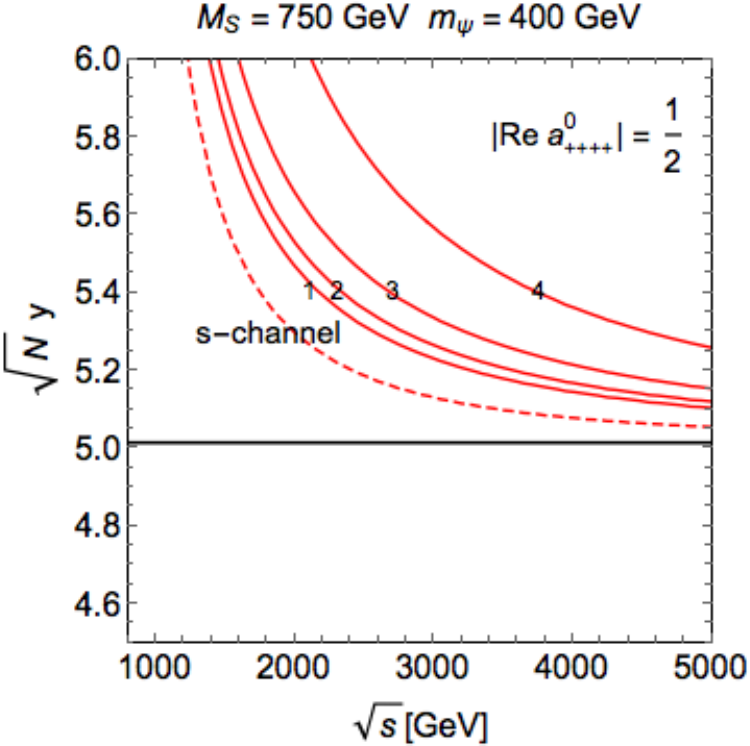}
  \end{center}
  \caption{\label{FSflav_yvssqrts_pppp}
  Tree-level unitarity bound in the $(\sqrt{s},\sqrt{N}y)$ plane for the reference values $M_S = 750$ GeV 
  and $m_\psi = 400$ GeV. 
  The dashed (red) line denotes the $s$-channel contribution (independent from $N$ in this normalization). 
  The full (red) lines, labelled by the value of $N=1,2,3,4$, denote instead the full contribution. 
  The value $y = \sqrt{8 \pi} \simeq 5$, indicated by the dashed (black) horizontal line, 
  is reached asymptotically.}
\end{figure}

The other non-zero scattering amplitude is the adjoint-adjoint one, which is found to be 
\begin{align}
\label{adjointchamp}
^{\ \ B}_\mathbf{\text{Adj}} \langle \psi \overline{\psi}  |S|\psi \overline{\psi}\rangle_\mathbf{\text{Adj}}^A 
&= T^{B\dag}_{kl} T^A_{ij} \langle \psi_k \overline{\psi}_l |S|\psi_i \overline{\psi}_j\rangle 
= T^{B}_{lk} T^A_{ij} (i \mathcal{T}_s \, \delta_{ij} \delta_{kl} + i \mathcal{T}_t \, \delta_{ik} \delta_{jl}) \nonumber \\
&= \Tr (T^{B}) \Tr (T^A) (i \mathcal{T}_s) +  \Tr (T^{B} T^A)  (i \mathcal{T}_t) 
= i \mathcal{T}_t \, \delta^{AB} \, .
\end{align}
Hence, we conclude that the adjoint-adjoint scattering is phenomenologically less relevant: 
only the subleading $t$-channel contributes, without the high-multiplicity enhancement. 

\item $y_{ij} = y_i \delta_{ij}$

This is the most general case relevant for a di-boson resonance, for which the mediators' couplings 
enter the partial width $\Gamma_{\gamma\gamma}$ as $\abs{\sum_i y_i}^2$. On the other hand, the unitarity 
bound on the $2\to 2$ scatterings applies to the combination $\sum_i \abs{y_i}^2$. 
Hence, at fixed value of $\abs{\sum_i y_i}^2$, the sum that enters in the amplitude for the $2 \to 2$ 
scattering is minimized when $y_i = y$ ($\forall$ $i$). In this way the bound from unitarity is minimized too. 

\end{enumerate}

Finally, we briefly discuss the case where the mediators carry extra gauge quantum numbers, 
as e.g.~color. This exactly matches the identity-$y$ scenario and thus all the previous results 
carry over. In particular, given an $N_R$-dimensional irreducible representation of the gauge group, 
the state corresponding to the gauge singlet combination always features an $N_R$ enhancement in the $s$-channel.

\subsection{Application to mediator models}

\begin{figure}[ht]
  \begin{center}
    \includegraphics[width=.32\textwidth]{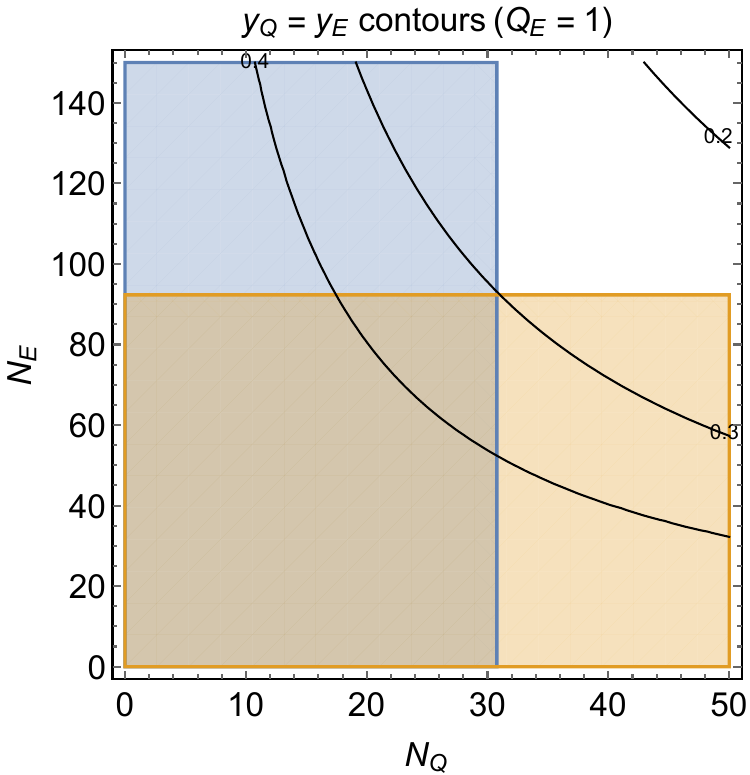}
    \includegraphics[width=.32\textwidth]{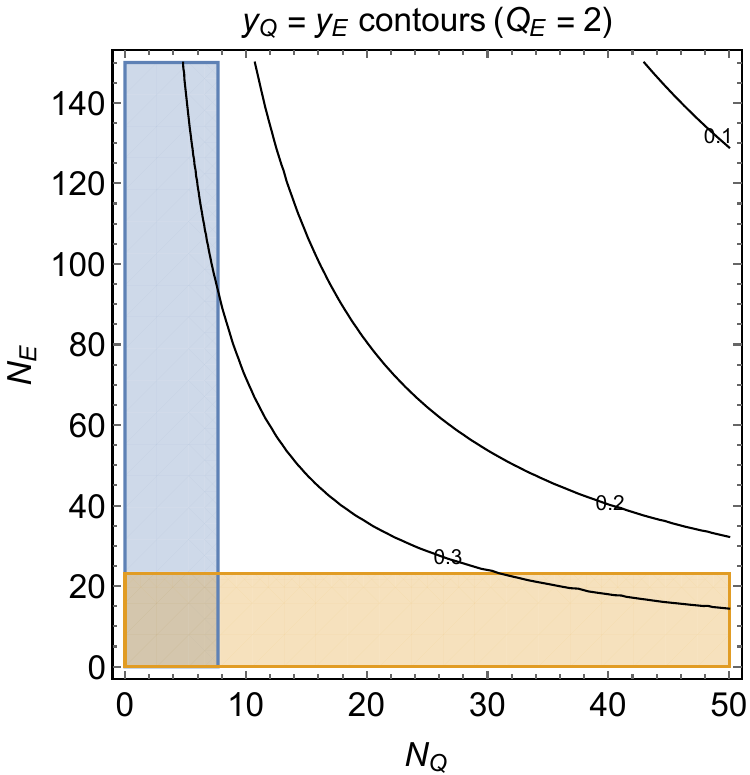}
    \includegraphics[width=.32\textwidth]{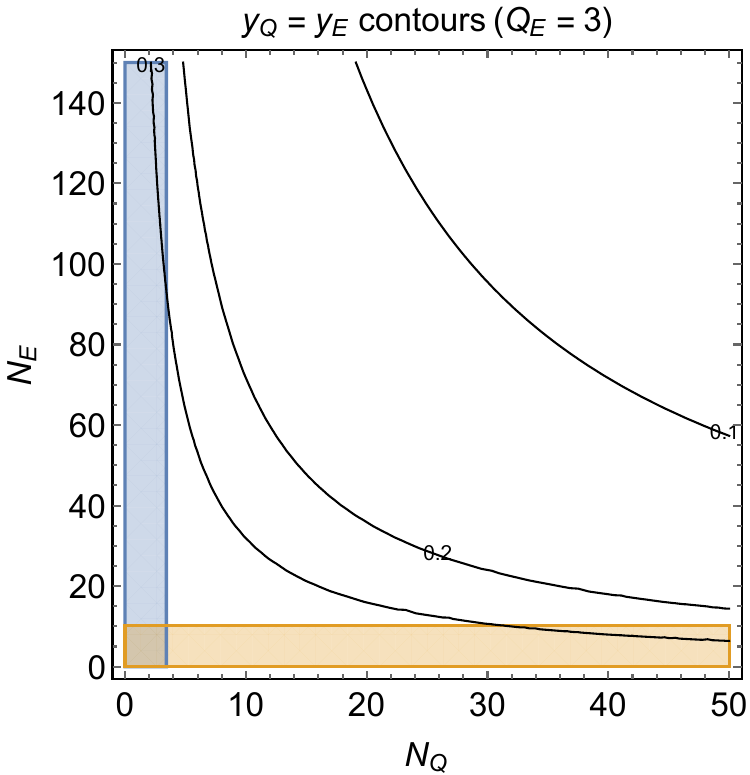}
    \includegraphics[width=.32\textwidth]{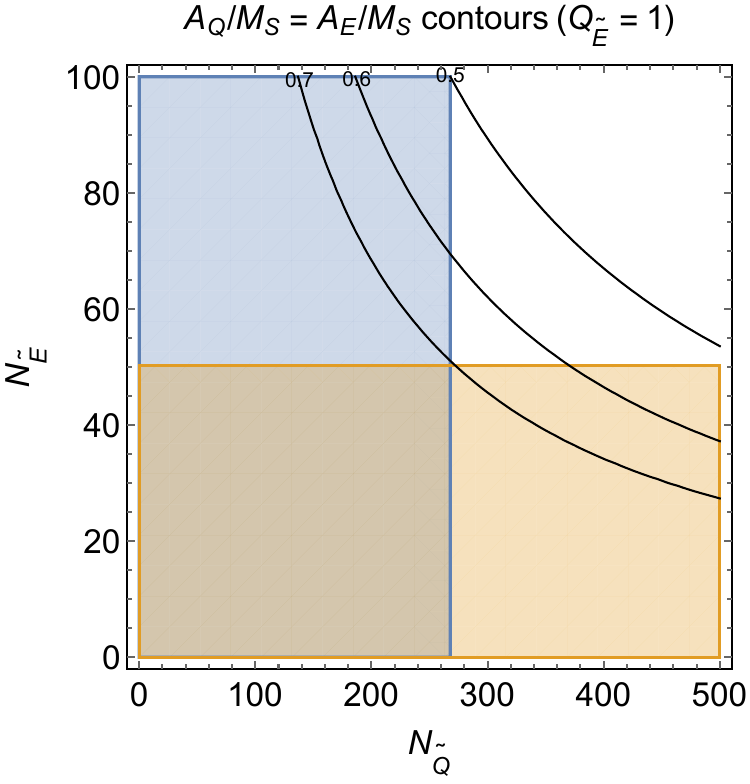}
    \includegraphics[width=.32\textwidth]{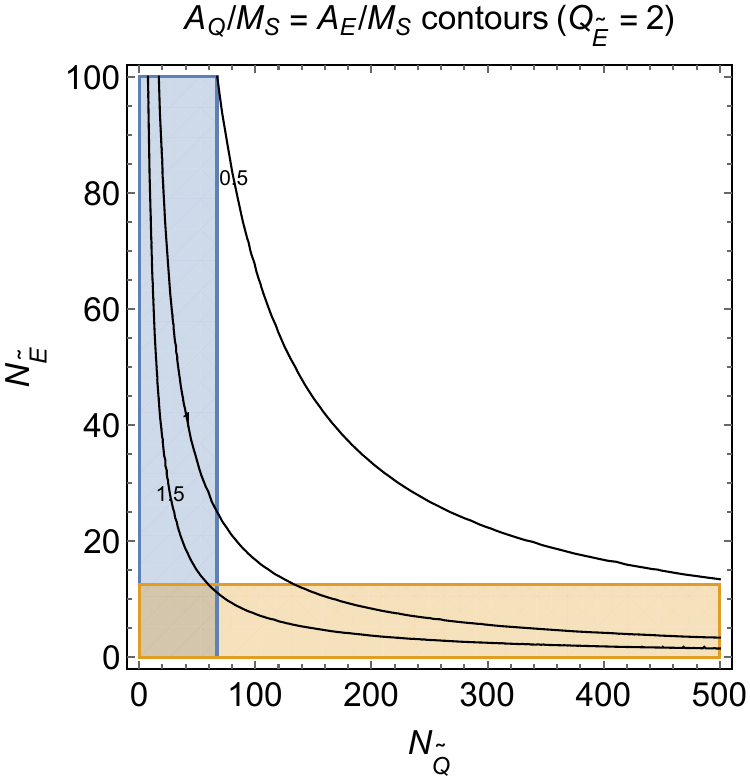}
    \includegraphics[width=.32\textwidth]{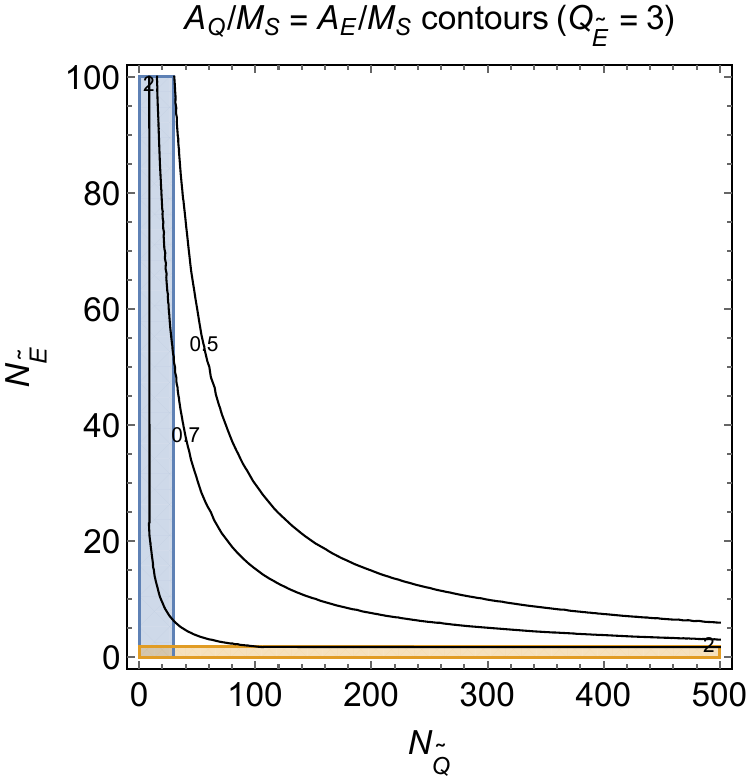}
  \end{center}
  \caption{\label{Excl_FSQ}
  Contours of constant Yukawa couplings $y_Q=y_E$ in the $N_Q$ vs.~$N_E$ plane (upper side plots) 
  and constant scalar trilinears $A_Q / M_S = A_E / M_S$ in the $N_{\tilde{Q}}$ vs.~$N_{\tilde{E}}$ plane (lower side plots) 
  for parameter points predicting a $\sigma_{\gamma\gamma}=1$~fb di-photon resonance with $M_S = 750$ GeV and $\Gamma_S / M_S = 0.1$ 
  (cf.~\eq{NFcond} and \eq{NScond}). 
  The different cases are associated to values of the EM charge of $Q_E$ and $Q_{\tilde{E}}$ from 1 to 3, 
  while the exclusion regions correspond to the tree-level unitarity bounds in \eqs{NEFcond}{NQFcond} (upper side plots) 
  and \eqs{NEScond}{NQScond} (lower side plots).}
\end{figure}

We are now ready to discuss the implication of the unitarity bounds 
on the required partial widths needed to reproduce any given $\gamma\gamma$ signal at the LHC. 
In particular, in the case of $gg$-initiated production processes (at $M_S=750$~GeV) the constraints to be fulfilled are the following: 
\begin{itemize}
\item Fermion mediators (model in \eq{LNF}, cf.~also \eq{unityuk}): 
\begin{align}
\label{NEFcond}
N_E y_E^2 &< \frac{8 \pi}{3} \, , \\
\label{NQFcond}
3 N_Q y_Q^2 &< \frac{8 \pi}{3} \, , \\ 
\label{NFcond} 
N^2_E N^2_Q y_E^2 y_Q^2 Q^4_E
&= 6.6 \times 10^4 \left(\frac{\sigma_{\gamma\gamma}}{\rm fb}\right)
\left( \frac{\Gamma_S/M_S}{0.1} \right) \, , 
\end{align}
The flavor and color enhancement of the bounds in \eqs{NEFcond}{NQFcond} 
hold in the asymptotic region $\sqrt{s} \gg M_S, m_{E,Q}$, 
where the partial wave is $s$-channel dominated, 
while in deriving \eq{NFcond} we used \eq{xsecgg} and \eq{GammaNF}.

\item Scalar mediators (model in \eq{LNS}): 
\begin{align}
\label{NEScond}
N_{\tilde{E}} \left( \frac{A_E}{M_S} \right)^2 & < 25 \, , 
\\
\label{NQScond}
3 N_{\tilde{Q}} \left( \frac{A_Q}{M_S} \right)^2 & < 400 \, , 
\\ 
\label{NScond}
N^2_{\tilde{E}} N^2_{\tilde{Q}}  
\left( \frac{A_E}{M_S} \right)^2 
\left( \frac{A_Q}{M_S} \right)^2 Q^4_{\tilde{E}}
&= 4.5 \times 10^7  \left(\frac{\sigma_{\gamma\gamma}}{\rm fb}\right)
\left( \frac{\Gamma_S/M_S}{0.1} \right) \, ,
\end{align}
The values in \eqs{NEScond}{NQScond} refer to the $s$-channel bounds 
of \fig{SS_AoMvssqrts}, for which the flavor and color enhancement apply, 
while in deriving \eq{NScond} we have used \eq{xsecgg} and \eq{GammaNS}. 
On the other hand, the following constraints (obtained by looking at the full partial wave amplitude in \fig{SS_AoMvssqrts})
\beq 
\left( \frac{A_E}{M_S} \right)^2 < 44 \, ,
\qquad 
\left( \frac{A_Q}{M_S} \right)^2 < 120 \, , 
\eeq
hold irrespectively of the flavor and color copies. Note that the bounds on $A_Q$ are weaker then on $A_E$ because the partial wave amplitudes 
are decreasing fast for heavy mediators (away from the poles). 
Thus, contrary to the fermionic case, unitarity bounds on these scalar couplings crucially depend on the assumed mediator masses. 
Nevertheless, the bounds cannot be circumvented by decoupling the mediator masses (for fixed $M_S$) 
since the decoupling of the partial rates in \eqs{PRgammagamma}{PRgg} is faster than that of the 
partial wave amplitude (cf.~\eq{a0scalar}).
\end{itemize}
In the case of fermion mediators we have 5 parameters ($y_E$, $y_Q$, $N_E$, $N_Q$ and $Q_E$) 
entering the expression in \eq{NFcond} corresponding to a particular di-photon signal strength. 
Hence, a possible way to display the tree-level unitarity bounds in \eqs{NEFcond}{NQFcond} is to 
choose a value of $Q_E$ and fix $y_Q=y_E$. 
\fig{Excl_FSQ} (upper side plots) displays iso-curves reproducing the benchmark signal of $\sigma_{\gamma\gamma}=1$~fb and $\Gamma_S / M_S = 0.1$ 
in the $N_Q$ vs.~$N_E$ plane and the 
associated perturbativity bounds for different values of $Q_E$. 
A very similar discussion applies to the case of scalar mediators (cf.~lower side plots). 

As it emerges from \fig{Excl_FSQ}, the only possibilities to accommodate the benchmark di-photon signal within weakly-coupled models 
are either via exotically-large EM charges\footnote{To this end, it would be relevant to consider scattering 
amplitudes providing unitarity constraints on the EM charge of the 
colorless mediators, e.g.~via hypercharge-mediated scatterings. 
However, unitarity arguments cannot be straightforwardly applied in presence of long-range forces, since the amplitudes 
are plagued by IR singularities (cf.~the case of Bhabha scattering in the forward 
region \cite{Peskin:1995ev}).} and/or a very large number of mediators' copies. 
These two latter options are also bounded by usual RGE 
arguments, which however are not sufficient to exclude such possibilities (see e.g.~\cite{Goertz:2015nkp}). 

\begin{figure}[!ht]
  \begin{center}
    \includegraphics[width=.40\textwidth]{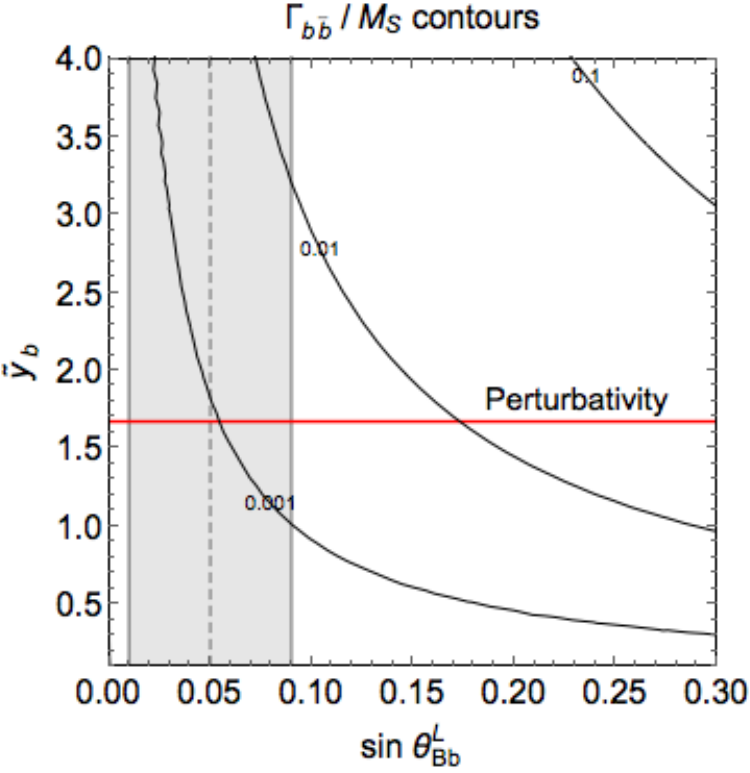}
  \end{center}
  \caption{\label{bbar_bounds}
  Contours of constant $\Gamma_{b\overline{b}}/M_S$ in the $({\sin\theta^L_{\mathcal{B}b}}, \tilde{y}_b)$ plane. 
  The values of $\Gamma_{b\overline{b}}/M_S$ are varied between $0.1$ and $0.001$. The vertical (grey) band denotes the 1-$\sigma$ upper bound on $\sin\theta^L_{\mathcal{B}b}$, while the full (red) line is the 
tree-level unitarity bound.  }
\end{figure}

We finally discuss the case of the model in \eq{VLmixing} where the production of $S$ is due to $b\overline{b}$-initiated processes.
Using \eq{xsecbbbar} and \eq{Sbb} we obtain 
\beq
\label{VLmixcond}
\left(\frac{\sin \theta^L_{\mathcal B b}}{0.05}\right)^2   \tilde y_b^2 = 77  \left(\frac{\sigma_{\gamma\gamma}}{\rm fb}\right) \left( \frac{\Gamma_S / M_S}{0.1} \right)  \left( \frac{\Gamma_{\gamma \gamma} / M_S}{10^{-4}} \right)^{-1} \, , 
\eeq
to be confronted with the tree-level unitarity bound 
\beq
\tilde{y}_b^2  <  \frac{8 \pi}{9} \, ,
\eeq
where we also took into account the color enhancement of the $s$-channel. 
In this case, the perturbative unitarity constraint is very severe (see \fig{bbar_bounds}). In particular for our benchmark it excludes the possibility for $S\to b\bar b$ decays to saturate a large decay width. 

\section{Conclusions}
\label{concl}

Perturbative unitarity is a powerful theoretical tool for inferring 
the range of validity of a given EFT, with notable examples of applications both in the physics 
of strong and electroweak interactions. 
The continued interest in di-boson resonances at the LHC motivated us to investigate the implications of partial wave unitarity for the theoretical description of such possible signals both in the minimal EFT extension of the SM as well as in its renormalizable UV completions.  

In the case of a TeV-scale scalar di-boson resonance observable at the LHC we have, under some very basic and natural assumptions on the structure of the EFT
(mainly that $S$ is a spin-0 SM gauge singlet and that the $\text{dim}=5$ operators in \eq{effLSM} are the most relevant ones for the decay of $S$), 
 demonstrated a potential violation of tree-level unitarity in the scattering of SM fields at energy scales of few tens of TeV.  
One should stress, however, that in many models 
(both weakly and strongly coupled) predicting observable di-boson resonances, 
new states are typically predicted to lie much below our energy estimates.

In a similar way one can use perturbative unitarity 
in order to estimate the range of validity of perturbation theory in explicit renormalizable UV completions of the low-energy EFT  
and accordingly set perturbativity bounds on the relevant model couplings. 
Especially in the case of a large total $S$ width, the inferred bounds are typically very constraining, 
and are in particular endangering the calculability of many weakly-coupled models present in the literature. 

Interestingly, tree-level unitarity bounds are important not only at high energies but also close to  thresholds of new physics. This is especially crucial for scalars interacting via relevant operators, since the corresponding unitarity bounds are always saturated at finite scattering energies relatively close to threshold. Other perturbativity criteria such as those based on Landau poles are only logarithmically sensitive to the energy scale and typically need a few decades of running before hitting the singularity of the Landau pole. 

Finally, we find that our perturbative bounds are sensitive not only to the strengths of the couplings ($y$) of the mediators to a di-boson resonance but also to the multiplicity $N$ of the mediator states.
For example, for fermions the bounds scale as $N y^2$, exhibiting a similar 't Hooft scaling as the perturbative bounds obtained by analyzing the RGE flow of the couplings~\cite{Goertz:2015nkp}. 


We conclude that in the event of an experimental observation of a scalar di-boson resonance at the LHC, while our estimates cannot provide a guarantee to see on-shell effects of additional new degrees of freedom at the LHC, they would immediately imply the existence of additional phenomena within the energy reach of the next generation 50-100~TeV hadron colliders, thus making a strong physics case for their construction. 


{
\section*{Note added} 
While completing this paper we came across Ref.~\cite{Cynolter:2016jxv}. Though part of our work overlaps with it, we reach different conclusions. 
}

\section*{Acknowledgments}

We thank Ramona Gr\"{o}ber, Jacobo L\'opez-Pav\'on, David Marzocca, Christopher W.~Murphy, Enrico Nardi, and Filippo Sala for helpful discussions. 
The work of L.D.L.~is supported by the Marie Curie CIG program, project number PCIG13-GA-2013-618439. J.F.K. acknowledges the financial support from the Slovenian Research Agency (research core funding No.\ P1-0035).

\appendix

\section{Amplitudes}
\label{Amplitudes}

In this Appendix we provide the details of the tree-level amplitude calculations.
We limit ourselves to the case of $2 \to 2$ scatterings in the center of mass frame and with all particle masses 
in the external states equal to $m$. By denoting the incoming momenta by $p$ and $k$ and the outgoing ones by $p'$ and $k'$, 
the kinematical variables are given by 
\begin{align}
p&=(E,0,0,p^3) \, , \\
k&=(E,0,0,-p^3) \, , \\
p'&=(E,p^3\sin\theta,0,p^3\cos\theta) \, , \\
k'&=(E,-p^3\sin\theta,0,-p^3\cos\theta) \, ,
\end{align}
with $p^3>0$. Correspondingly, the Mandelstam variables read
\begin{align}
s &= (p+k)^2 = 4E^2 \, , \\
t &= (p-p')^2 = -4(p^3)^2\sin^2\frac{\theta}{2} \, , \\
u &= (p-k')^2 = -4(p^3)^2\cos^2\frac{\theta}{2} \, ,
\end{align}
Everything can be conveniently re-expressed in terms of $\sqrt{s}$ and $m$ via the relations
$E=\sqrt{s}/2$, $(p^3)^2 = s/4 - m^2$, $t = - (s - 4 m^2) \sin^2\frac{\theta}{2}$ and 
$u = - (s - 4 m^2) \cos^2\frac{\theta}{2}$.

\subsection{$\gamma \gamma \to \gamma \gamma$ scattering}
\label{EFTscattering}

Given the interaction Lagrangian 
\beq
\mathcal{L}_I \supset c \, S F^{\mu \nu} F_{\mu \nu} \, , 
\eeq
we are interested in computing the scattering amplitude for the process 
$\gamma(p,s_1,\mu) + \gamma(k,s_2,\nu) \to \gamma(p',s_3,\alpha) + \gamma(k',s_4,\beta)$. 
To this end, it is convenient to choose a specific basis for the transverse polarization vectors 
\begin{align*}
\epsilon_+ (p) & =  \frac{1}{\sqrt{2}} (0,1,-i,0) \, , & \epsilon_-  (p) & =  \frac{1}{\sqrt{2}} (0,1,i,0) \, , \\
\epsilon_+ (k) & = - \epsilon_- (p) \, , & \epsilon_- (k) & = - \epsilon_+ (p) \, , \\
\epsilon_+ (p') & = \frac{1}{\sqrt{2}} (0,\cos \theta,-i, - \sin \theta) \, , & \epsilon_-  (p') & =  \frac{1}{\sqrt{2}} (0,\cos \theta,i, -\sin \theta) \, , \\
\epsilon_+ (k') & = - \epsilon_- (p') \, , & \epsilon_- (k') & = - \epsilon_+ (p') \, .  
\end{align*} 
Since we focus our analysis on states with $J=0$, 
we can restrict ourselves to $s \equiv s_1=s_2$ and $r \equiv s_3=s_4$. 
In such a case all the amplitudes are proportional to 
\begin{equation}
 \epsilon_{s \mu} (p) \epsilon_{s \nu} (k)  \epsilon^*_{r \alpha} (p') \epsilon^*_{r \beta} (k') =
 \epsilon_{s \mu} (p) \epsilon^*_{s \nu} (p)  \epsilon^*_{r \alpha} (p') \epsilon_{r \beta} (p') \, .
\end{equation}
For later convenience, let us also define $\epsilon_{s \mu} \equiv \epsilon_{s \mu} (p)$ 
and $\epsilon_{r \mu} \equiv \epsilon_{r \mu} (p')$. 
We then get the following contributions for the amplitude in the $s$, $t$ and $u$ channels
\begin{align}
 \mathcal{T}_s &=  \frac{- 16  c^2}{s -m_S^2} \epsilon_{s \mu} \epsilon^*_{s \nu} \epsilon^*_{r \alpha} \epsilon_{r \beta} \left[ (p \cdot k) g^{\mu \nu} - p^{\nu} k^{\mu}  \right]
[ (p' \cdot k') g^{\mu \nu} - p'^{\beta} k'^{\alpha} ] \nonumber \\
&= \frac{- 16  c^2}{s -m_S^2} (\epsilon_{s} \cdot \epsilon^*_{s})  (\epsilon^*_{r} \cdot \epsilon_{r}) (p \cdot k) (p' \cdot k') = -4  c^2 \frac{s^2}{s-m_S^2} \, , \\
 \mathcal{T}_t &=  \frac{- 16  c^2}{t -m_S^2} \epsilon_{s \mu} \epsilon^*_{s \nu} \epsilon^*_{r \alpha} \epsilon_{r \beta} \left[ (-p \cdot p') g^{\mu \alpha} + p^{\alpha} p'^{\mu}  \right]
[ (-k \cdot k') g^{\nu \beta} + k^{\beta} k'^{\nu} ] \nonumber \\
&= \frac{- 16  c^2}{t -m_S^2} \left| -(p \cdot p') (\epsilon_s \cdot \epsilon_r^*) + (p \cdot \epsilon^*_r) (p' \cdot \epsilon_s)  \right|^2 = -4  c^2 \frac{t^2}{t-m_S^2} \delta_{s,-r}
\, , \\
 \mathcal{T}_u &=  \frac{- 16  c^2}{u -m_S^2} \epsilon_{s \mu} \epsilon^*_{s \nu} \epsilon^*_{r \alpha} \epsilon_{r \beta} \left[ (-p \cdot k') g^{\mu \beta} + p^{\beta} k'^{\mu}  \right]
[ (-k \cdot p') g^{\nu \alpha} + k^{\alpha} p'^{\nu} ] \nonumber \\
&= \frac{- 16  c^2}{u -m_S^2} \left| -(p \cdot k') (\epsilon_s \cdot \epsilon_r) + (p \cdot \epsilon^*_r) (k' \cdot \epsilon^*_s)  \right|^2= -4  c^2 \frac{u^2}{u-m_S^2} \delta_{s,-r} \, .
\end{align}

\subsection{$\psi \overline{\psi} \to \psi \overline{\psi}$ scattering}
\label{psipsibarpsipsibar}

Starting from the interaction Lagrangian in \eq{intSpsibarpsi}, 
the $s$- and $t$-channel scattering amplitudes 
for the process $\psi(p,r) + \overline{\psi}(k,s) \to \psi(p',r') + \overline{\psi}(k',s')$ are
\begin{align}
\label{M1sch}
\mathcal{T}_s &= - \frac{y^2}{s-M_S^2} \overline{v}^{s} (k) u^{r} (p) \overline{u}^{r'} (p') v^{s'} (k') \, , \\
\label{M1tch}
\mathcal{T}_t &= + \frac{y^2}{t-M_S^2} \overline{u}^{r'} (p') u^{r} (p) \overline{v}^{s} (k) v^{s'} (k') \, .
\end{align}
To evaluate the amplitudes we consider the general representation for the spinor 
polarizations (see e.g.~\cite{Peskin:1995ev})
\beq
u^{r} (p) = 
\left( 
\begin{array}{c}
\sqrt{p \cdot \sigma} \, \xi_r \\
\sqrt{p \cdot \bar\sigma} \, \xi_r
\end{array}
\right) \, , \qquad
v^{s} (p) = 
\left( 
\begin{array}{c}
\sqrt{p \cdot \sigma} \, \eta_s \\
-\sqrt{p \cdot \bar\sigma} \, \eta_s
\end{array}
\right) \, ,
\eeq
for $r= +, -$ and $s= +, -$.  In particular, 
$\sigma = (1,\vec{\sigma})$ and $\bar\sigma = (1,-\vec{\sigma})$, where 
$\vec{\sigma} = (\sigma^1,\sigma^2,\sigma^3)$ 
is the vector Pauli matrix, while $\xi_r$ and $\eta_s$ 
provide two independent bases for  
two-component spinors. 
The latter are chosen according to the following convention on the definition of the spinors' helicities
\beq
\label{helicity}
\left(\vec\Sigma \cdot \hat{p}\right) u^r (p) = r u^r (p) \, , \qquad
\left(\vec\Sigma \cdot \hat{p} \right) v^s (p) = - s v^s (p) \, ,
\eeq 
where $\vec\Sigma = \text{diag}(\vec\sigma,\vec\sigma)$ denotes the spin operator. 
Note that for anti-particles the helicity is defined with the opposite sign. 
A standard basis, for the two component spinors, which satisfies \eq{helicity} is provided by 
\beq
\xi_+ = 
\left( 
\begin{array}{c}
1 \\
0
\end{array}
\right) \, ,\qquad
\xi_- = 
\left( 
\begin{array}{c}
0 \\
1
\end{array}
\right) \, ,\qquad
\eta_+ = 
\left( 
\begin{array}{c}
0 \\
1
\end{array}
\right) \, ,\qquad
\eta_- = 
\left( 
\begin{array}{c}
1 \\
0
\end{array}
\right) \, .
\eeq
In order to evaluate \eqs{M1sch}{M1tch}, we need the rotated spinors
\beq
u(p') = 
\left( 
\begin{array}{cc}
R(\theta) & 0 \\
0 & R(\theta) 
\end{array}
\right) u(p) 
= 
\left( 
\begin{array}{cc}
\underbrace{R(\theta)  \sqrt{p \cdot \sigma}  R(\theta)^{-1}}_{\sqrt{p' \cdot \sigma} } & 0 \\
0 & \underbrace{R(\theta) \sqrt{p \cdot \bar\sigma} R(\theta)^{-1}}_{\sqrt{p' \cdot \bar\sigma}}
\end{array}
\right)
\left( 
\begin{array}{c}
R(\theta) \xi_r \\
R(\theta) \xi_r
\end{array}
\right) \, ,
\eeq
where 
\beq 
R(\theta) = 
\left( 
\begin{array}{cc}
\cos\frac{\theta}{2} & -\sin\frac{\theta}{2} \\
\sin\frac{\theta}{2} & \cos\frac{\theta}{2}
\end{array}
\right) \, ,
\eeq 
is the rotation matrix of a bi-spinor 
in the 1-3 plane by an angle $\theta$ with respect to the 2nd axis. 
$R(\pi)$ and $R(\theta+\pi)$ are instead the relevant rotation matrices for $v(k)$ and $v(k')$, respectively. 
The helicity amplitudes for the $\psi \overline{\psi} \to \psi \overline{\psi}$ scattering (\eqs{M1sch}{M1tch})
are displayed in \Table{FFbar}. 

\begin{table}[htbp]
\begin{footnotesize}
\renewcommand{\arraystretch}{1.8}
\centering
\begin{tabular}{@{} |c|c|c|c|c|c|c|c| @{}}
\hline
$r$ &  $s$ & $r'$ & $s'$ & $\overline{v}^{s} (k) u^{r} (p) \overline{u}^{r'} (p') v^{s'} (k')$ & $\overline{u}^{r'} (p') u^{r} (p) \overline{v}^{s} (k) v^{s'} (k')$ 
& $\mathcal{T}_s (E\gg)$ & $\mathcal{T}_t (E\gg)$ \\
\hline
\hline
$+$ & $+$ & $+$ & $+$ & $4 (p^3)^2$ & $- 4 m^2 \cos^2\frac{\theta}{2}$ & $-y^2$ & $0$ \\
$+$ & $-$ & $+$ & $+$ & $0$ & $2 m E \sin{\theta}$ & $0$ & $0$  \\
$-$ & $+$ & $+$ & $+$ & $0$ & $- 2 m E \sin{\theta}$ & $0$ & $0$  \\
$-$ & $-$ & $+$ & $+$ & $ 4 (p^3)^2$ & $4 E^2 \sin^2\frac{\theta}{2}$ & 
$-y^2$ & $-y^2$  \\
$+$ & $+$ & $+$ & $-$ & $0$ & $- 2 m E \sin{\theta}$ & $0$ & $0$ \\
$+$ & $-$ & $+$ & $-$ & $0$ & $- 4 m^2 \cos^2\frac{\theta}{2}$ & $0$ & $0$  \\
$-$ & $+$ & $+$ & $-$ & $0$ & $- 4 E^2 \sin^2\frac{\theta}{2}$ & $0$ & $y^2$  \\
$-$ & $-$ & $+$ & $-$ & $0$ & $- 2 m E \sin{\theta}$ & $0$ & $0$  \\
$+$ & $+$ & $-$ & $+$ & $0$ & $ 2 m E \sin{\theta}$ & $0$ & $0$ \\
$+$ & $-$ & $-$ & $+$ & $0$ & $- 4 E^2 \sin^2\frac{\theta}{2}$ & $0$ & $y^2$  \\
$-$ & $+$ & $-$ & $+$ & $0$ & $- 4 m^2 \cos^2\frac{\theta}{2}$ & $0$ & $0$  \\
$-$ & $-$ & $-$ & $+$ & $0$ & $ 2 m E \sin{\theta}$ & $0$ & $0$  \\
$+$ & $+$ & $-$ & $-$ & $ 4 (p^3)^2$ & $ 4 E^2 \sin^2\frac{\theta}{2}$ & 
$-y^2$ & $-y^2$ \\
$+$ & $-$ & $-$ & $-$ & $0$ & $ 2 m E \sin{\theta}$ & $0$ & $0$  \\
$-$ & $+$ & $-$ & $-$ & $0$ & $- 2 m E \sin{\theta}$ & $0$ & $0$  \\
$-$ & $-$ & $-$ & $-$ & $4 (p^3)^2$ & $- 4 m^2 \cos^2\frac{\theta}{2}$ & 
$-y^2$ & $0$  \\
  \hline
  \end{tabular}
  \caption{\label{FFbar} 
  $\psi \overline{\psi} \to \psi \overline{\psi}$
    helicity amplitudes
    for the interaction term in \eq{intSpsibarpsi}.}
    \end{footnotesize}
\end{table}


Finally, we report for completeness the analytical expression of the $J=0$ partial wave 
for the $++++$ and $++--$ helicity-state scattering amplitudes, which read respectively 
\begin{align}
\label{a0++++full}
a^0_{++++}&= - y^2 \frac{\sqrt{s(s-4m^2_\psi)}}{16\pi s} \nonumber \\
&\times \left( \frac{s-4m^2_\psi}{s-M^2_S} 
- \frac{4 m^2_\psi \left( \left( s-4m^2_\psi + M^2_S \right) 
\log\frac{s-4m^2_\psi + M^2_S}{M^2_S} - \left( s-4m^2_\psi \right) 
\right)}{\left( s-4m^2_\psi \right)^2} \right)
\, , \\
\label{a0++--full}
a^0_{++--}&= - y^2 \frac{\sqrt{s(s-4m^2_\psi)}}{16\pi s} 
\left( \frac{s-4m^2_\psi}{s-M^2_S} 
- \frac{s \left( M^2_S
\log\frac{s-4m^2_\psi + M^2_S}{M^2_S} - \left( s-4m^2_\psi \right)
\right)}{\left( s-4m^2_\psi \right)^2} \right)
\, .
\end{align}

\clearpage

\bibliographystyle{utphys.bst}
\bibliography{bibliography}

\end{document}